\documentclass[twocolumn,aps,superscriptaddress]{revtex4-1}
\usepackage[dvipsnames]{xcolor}
\usepackage{tikz}
\usetikzlibrary{arrows}
\usetikzlibrary{decorations.markings}
\usetikzlibrary{positioning}
\usepackage{siunitx}
\usepackage{standalone}
\usepackage{amsmath}
\usepackage{amsfonts}
\usepackage{amssymb}
\usepackage{hyperref}
\usepackage{graphicx}
\usepackage{bm}
\newcommand{\heaviside}{\Theta}

\begin{document}
\title{Symmetry breaking in sticky collisions between ultracold molecules}

\author{Marijn P. Man}
\affiliation{Institute for Molecules and Materials, Radboud University, Nijmegen, The Netherlands}
\author{Gerrit C. Groenenboom}
\affiliation{Institute for Molecules and Materials, Radboud University, Nijmegen, The Netherlands}
\author{Tijs Karman}
\email{tkarman@science.ru.nl}
\affiliation{Institute for Molecules and Materials, Radboud University, Nijmegen, The Netherlands}

\begin{abstract}
Ultracold molecules undergo ``sticky collisions'' that result in loss even for chemically nonreactive molecules.
Sticking times can be enhanced by orders of magnitude by interactions that lead to non-conservation of nuclear spin or total angular momentum.
We present a quantitative theory of the required strength of such symmetry-breaking interactions based on classical simulation of collision complexes.
We find static electric fields as small as $10$~V/cm can lead to non-conservation of angular momentum,
while we find nuclear spin is conserved during collisions.
We also compute loss of collision complexes due to spontaneous emission and absorption of black-body radiation, which are found to be slow.
\end{abstract}

\maketitle

Ultracold molecules are promising for the realization of quantum simulation with tunable anisotropic long-range interactions \cite{micheli:06,buchler:07,cooper:09,carr:09},
quantum computation \cite{demille:02, yelin:06,ni:18,kaufman:21},
precision measurement \cite{andreev:18,ho:20},
and the exploration of chemistry in a coherent quantum mechanical regime \cite{krems:08, ospelkaus:10,liu:21,hu:21,liu:22}.
Unfortunately, the realization of these prospects has been hampered by universal collisional losses that occur in short-range encounters between molecules.
These collisional losses limit the lifetime of ultracold molecular gases~\cite{ospelkaus:10,takekoshi:14, molony:14,guo:16,park:15},
limit our ability to collisionally cool molecules to high phase-space densities~\cite{son:20,valtolina:20,matsuda:20,li:21,schindewolf:22},
and spoil Feshbach resonances that could otherwise form a control knob to tune interactions~\cite{yang:19,yang:22,son:22,su:22}.
Initially, collisional losses were attributed to chemical reactions~\cite{ospelkaus:10,idziaszek:10},
but similar loss has been observed for non-reactive species~\cite{takekoshi:14, molony:14,guo:16,park:15},
and therefore it is not fully understood what causes or how one might eliminate collisional loss.

Pioneering work by Mayle \emph{et al.}~\cite{mayle:12,mayle:13} has proposed that collisions between ultracold molecules might be ``sticky''.
Collision complexes have a high density of states, yet at ultracold temperatures can only dissociate in a single scattering channel.
The idea put forward is that the collision complexes will chaotically explore available phase space,
leading to excessively long sticking times determined by the density of states through Rice-Ramsperger-Kassel-Marcus (RRKM) theory.
During these sticky collisions the molecules are vulnerable to collisions with a third body~\cite{mayle:12,mayle:13,christianen:19b} or photoexcitation by the trapping laser~\cite{christianen:19a}.
This work has shaped the way the field thinks about collisional loss \cite{croft:20,croft:21,quemener:22,christianen:21}, although the debate is not settled on even the order of magnitudes of sticking times \cite{nichols:22}, nor on what physical loss processes might occur during that time.

Some of the present authors have proposed a theoretical framework to compute the density of states of ultracold collision complexes \cite{christianen:19b},
and the rate of loss by photoexcitation of these complexes by the trapping laser \cite{christianen:19a}.
These predictions were later confirmed quantitatively by two independent experiments using reactive KRb molecules \cite{liu:20} and nonreactive RbCs \cite{gregory:20}.
These experiments used modulated optical dipole traps,
which maintain a time-averaged trapping potential while the dark time, during which the modulated trap is off, can exceed the collision complex' sticking time.
This reduced photoexcitation loss even though the loss rate is saturated with light intensity when the trap light is on.
Subsequent experiments on NaK and NaRb molecules, however, could not observe such suppression of collisional loss in a modulated trap \cite{gersema:21},
nor could collisional loss be eliminated by trapping the molecules in a repulsive box potential \cite{bause:21}.
These observations remain unexplained, and form the main motivation for the present study.

It has been speculated that the sticking time could be enhanced by orders of magnitude by symmetry breaking that leads to the non-conservation of otherwise conserved quantum numbers,
as this would drastically increase the effective phase-space volume that collision complexes explore ergodically.
In particular, total angular momentum is strictly conserved \emph{only} in the absence of external fields.
Using the methods of Ref.~\cite{christianen:19b} one can compute sticking times in the two limiting cases that total angular momentum is either strictly conserved or completely scrambled,
but it is unclear at what external field strengths the transition between these limits occurs.
Similarly, there are hints that nuclear spin degrees of freedom do not participate in the collision dynamics \cite{hu:21},
but it is unclear for which molecules and under which conditions this is the case.
Another possible explanation is that there exist additional loss mechanisms that limit the lifetime of collision complexes in the dark.
It is, however, difficult to conclusively rule out \emph{all possible} loss mechanisms.

In this letter, we develop classical simulations of ultracold collision complexes.
First, we show the simulated dynamics is consistent with chaotic dynamics and RRKM theory.
Next, we study the effect of a finite electric field.
Due to the dynamics of a collision complex, its dipole moment fluctuates rapidly.
The autocorrelation function of these fluctuations determines the coupling between different total-angular-momentum states.
We find static electric fields as small as $10$~V/cm can lead to non-conservation of angular momentum.
We also apply the formalism to nuclear spin couplings,
and find nuclear spin states are conserved during the sticking time in NaK+NaK and RbCs+RbCs collisions.
Finally, we apply the dipole autocorrelation function to compute loss of collision complexes by spontaneous emission and absorption of black-body radiation.
Both processes result in slow loss.

\begin{figure}
\centering
\includegraphics[width=0.475\textwidth]{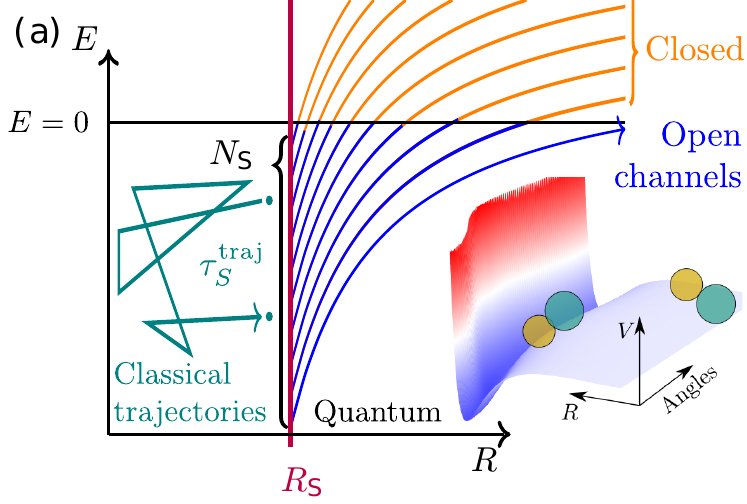}
\includegraphics[width=0.475\textwidth]{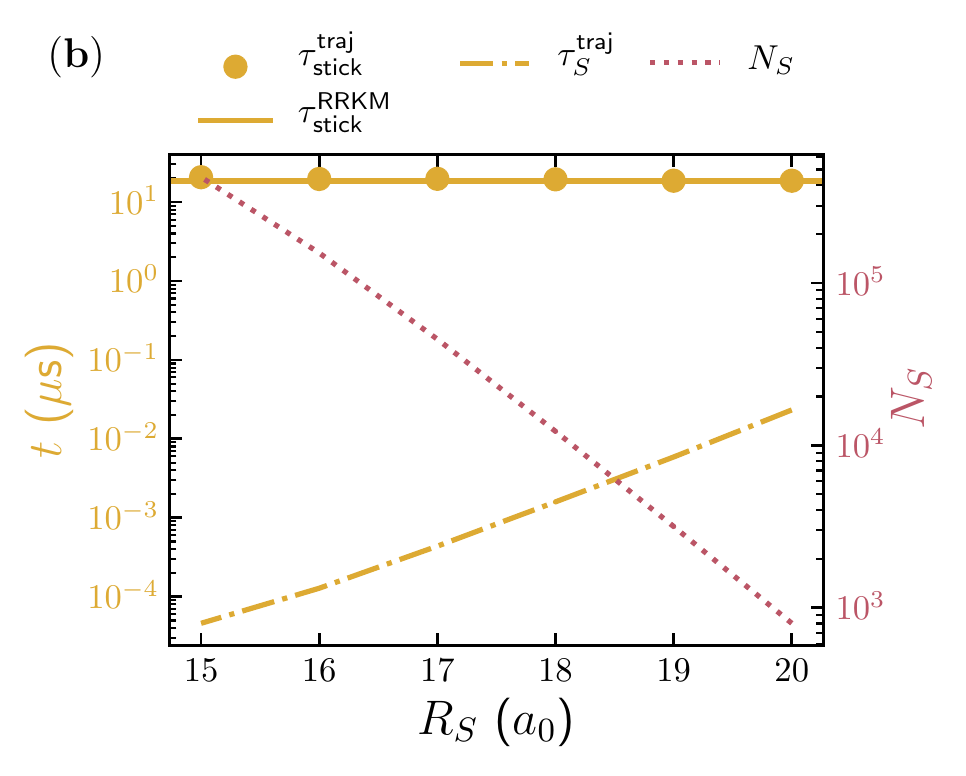}
\caption{
{\bf Classical simulations of ultracold sticky collisions.}
Panel ({\bf a}) shows schematically the classical simulations on a simple but realistic high-dimensional potential energy surface. 
After a typical time, $\tau_S^\mathrm{traj}$, the classical trajectory crosses the dividing surface $R=R_S$.
At this surface, $N_S \gg 1$ collision channels are energetically accessible.
We estimate the physical sticking time in ultracold collisions,
where only a single collision channel is open,
by correcting as $\tau_\mathrm{stick}^\mathrm{traj} = N_S \tau_S^\mathrm{traj}$.
Panel ({\bf b}) demonstrates quantitatively the steep dependence of $\tau_S^\mathrm{traj}$ and $N_S$ on the position of the dividing surface, $R_S$,
and that the sticking time estimated from classical trajectories, $\tau_\mathrm{stick}^\mathrm{traj}$, is independent of $R_S$ and in quantitative agreement with sticking times obtained from phase-space integrals, $\tau^\mathrm{RRKM}_\mathrm{stick}$.
}
\label{fig:simulation}
\end{figure}

We perform classical simulations of the sticky ultracold collision complexes NaK+K, NaK+NaK, and RbCs+RbCs, as illustrated in Fig.~\ref{fig:simulation}(a) and explained in detail in the Supplement \cite{supp}.
In short, we model the interactions using diatomics-in-molecules \cite{ellison:63,supp},
which describes spin-dependent pairwise interactions between all atoms.
The trajectories were initialized with molecules at their equilibrium bond length, for random orientations, and without vibrational or rotational kinetic energy, ensuring zero total angular momentum.
The radial kinetic energy was chosen such that the total energy equaled the lowest dissociation limit.
After a short thermalization time, we then simulated classical trajectories using a fourth-order symplectic propagator,
which ensures conservation of phase-space volume and long-term numerical stability.

First, we investigate whether the simulated dynamics of the collision complexes is consistent with RRKM theory,
as has been proposed by Mayle \emph{et al.}~\cite{mayle:12,mayle:13}.
The central result in RRKM theory is that the time spent in a phase-space region is given by
\begin{align}
\tau^\mathrm{RRKM}_S = \frac{2\pi \hbar \rho_S}{N_S},
\label{eq:rrkm}
\end{align}
where $\rho_S$ is the density of states in that region, and $N_S$ is the number of states at the boundary, which is known as the dividing surface.
Applied to ultracold collisions where at dissociation there is only a single open channel, $N_S=1$, this yields the sticking time $\tau_\mathrm{stick}^\mathrm{RRKM} = 2\pi\hbar\rho_S$.
Furthermore, there exists a clear separation of length scales of the complex, where the density of states $\rho_S$ is supported by intermolecular separations $R$ shorter than tens of bohr radii \cite{supp},
and the length scale of long-range interactions of hundreds of bohr radii.
Hence, one can unambiguously define a suitable dividing surface at intermediate $R$.

When simulating the dynamics classically, however,
at zero collision energy $N_S\rightarrow 0$, not $N_S\rightarrow 1$, as the dividing surface is moved outwards.
Hence, one cannot converge the calculation by moving the dividing surface outwards.
Instead, we pick a dividing surface at a convenient distance $R_S \simeq 20$~$a_0$.
We then observe that the mean time elapsed before the trajectory crosses the dividing surface, $\tau_S^\mathrm{traj}$, becomes consistent with Eq.~\eqref{eq:rrkm} if we account for the larger number of states $N_S$ at the dividing surface, which is computed independently as a phase space integral.
By extrapolating to the physical case where $N=1$, we recover sticking times $\tau_\mathrm{stick}^\mathrm{traj} = N_S \tau_S^\mathrm{traj}$, in agreement with RRKM predictions based on density of states computed independently using phase-space integrals, see Fig.~\ref{fig:simulation}(b).
This analysis provides support for the physical picture proposed by Mayle \emph{et al.} \cite{mayle:12,mayle:13} of classically chaotic dynamics, as subsequently assumed in further work \cite{croft:20,croft:21,quemener:22,christianen:21}.
Our approach has the remarkable advantage that we can study sticking times that are orders of magnitude larger than the time simulated.

Based on the above analysis, we caution sticking times cannot be extracted directly from classical simulations of sticking dynamics without accounting for the effective number of states at the dividing surface~\cite{croft:14,klos:21},
which is furthermore impacted by inclusion of a finite collision energy, zero-point energy, or the numerically imperfect conservation of energy \cite{supp}.

We now turn our attention to the non-conservation of quantum numbers by a symmetry-breaking perturbation.
Following Feingold and Peres \cite{feingold:86}, transition moments of $\hat{A}$ between energy-eigenstates in a classically chaotic system can be related to fluctuations of the classical observable, $A(t)$, in a micro-canonical ensemble.
These fluctuations are quantified by the autocorrelation function (ACF) and the coupling between energy eigenstates is given by
\begin{align}
|\langle i | \hat{A} | j \rangle|^2 = \frac{S_A(E,\omega_{ij})}{\hbar \rho},
\end{align}
where 
\begin{align}
S_A(E,\omega) = \frac{1}{2\pi} \int_{-\infty}^\infty \langle A(0) A(t') \rangle\ \exp(i\omega t')\ dt'
\end{align}
is the Fourier transform of the coupling's ACF that is computed classically, $\langle \cdots \rangle$ indicates a micro-canonical ensemble average, and $\hbar\omega_{ij}$ is the energy difference between energy level $i$ and $j$.
We use the static limit $S_A(E,\omega) \approx S_A(E,0)$ because the typical transition frequency, in the order of $1/h\rho$, is much smaller than $1/\tau_\mathrm{ACF}$,
where $\tau_\mathrm{ACF}$ is a typical timescale for the decay of the ACF.
The static limit $S_A(E,0)$ is simply the time integral of the ACF, and roughly in the order of $\langle A^2 \rangle \tau_\mathrm{ACF}$.
The mean magnitude of the coupling between energy eigenstates is then in the order of $A \sqrt{\tau_\mathrm{ACF} / \tau_\mathrm{stick}}$,
which is handwavingly interpreted as the magnitude of the bare coupling, $A$, that is dynamically reduced if fluctuations of $A$ are fast compared to the sticking time.

\begin{figure}
\centering
\includegraphics[width=0.475\textwidth]{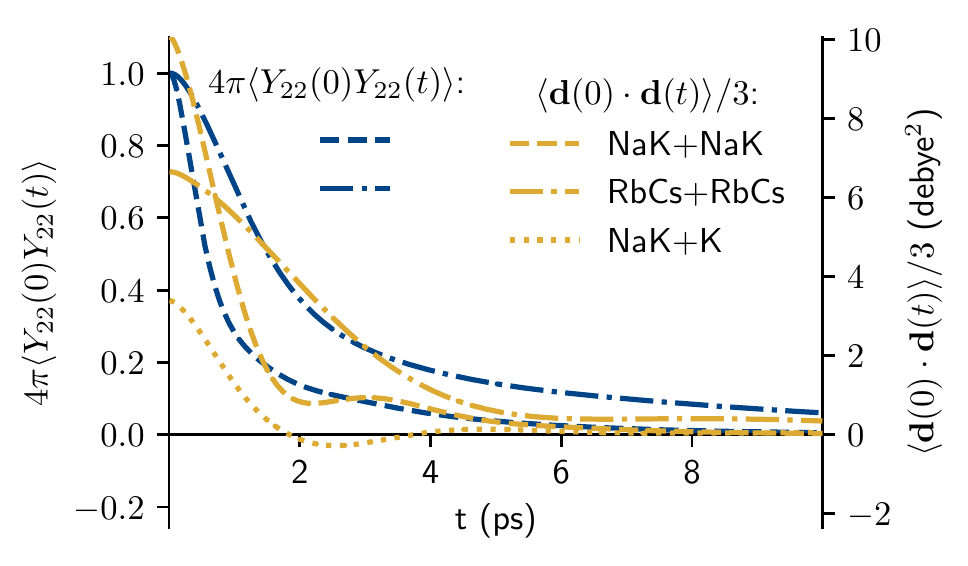}
\caption{
{\bf Autocorrelation functions} (ACFs).
Dipole and spherical-harmonic ACFs for NaK+K, NaK+NaK, and RbCs+RbCs sticky collisions.
}
\label{fig:ACF}
\end{figure}

We specialize the discussion to coupling between $J$ and $J'=J\pm 1$ states by the Stark interaction $\hat{A} = - \hat{\bm{d}} \cdot \bm{E}$ between the complex' dipole moment, $\bm{d}$, and an external static electric field, $\bm{E}$.
We simulated dipole ACFs for NaK+K, NaK+NaK, and RbCs+RbCs collision complexes in zero field, see the Supplement for details \cite{supp}.
The resulting ACFs are shown in Fig.~\ref{fig:ACF}.

\begin{figure}
\centering
\includegraphics[width=0.5\textwidth]{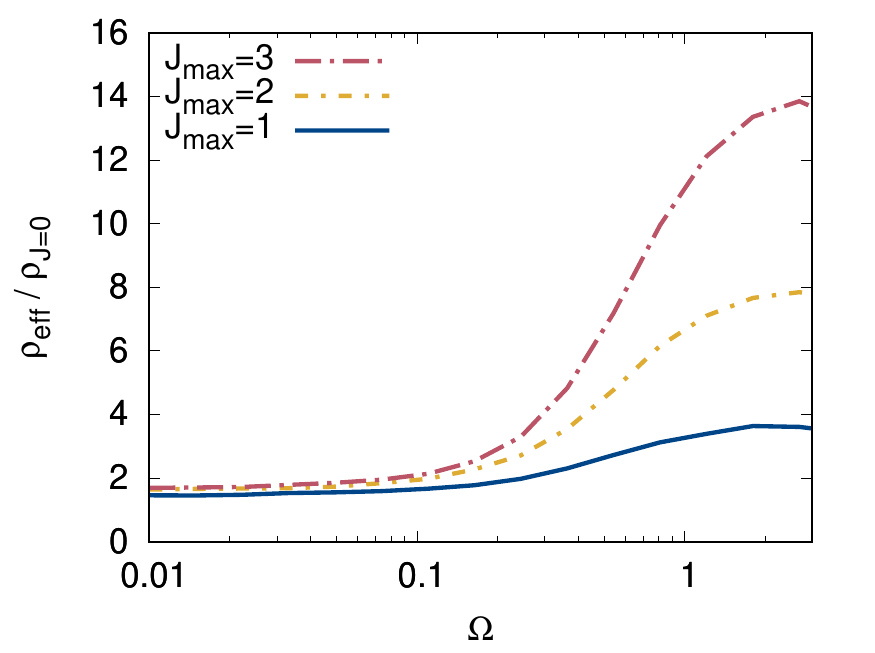}
\caption{
{\bf Effective density of states} as determined by fitting the distribution of sticking times, see Supplement~\cite{supp}.
The dimensionless parameter $\Omega$ represents the Stark interaction in units of the level spacing, and hence effectively represents the electric field.
The total density of states increases as $(1+J_\mathrm{max})^2$ for large $\Omega$, but the transition to $J$-conservation consistently occurs around $\Omega=1$ irrespective of the value of $J_\mathrm{max}$.
}
\label{fig:Omega}
\end{figure}

Following Leitner \emph{et al.}~\cite{leitner:94}, we characterize the transition from $J$ conservation to non-conservation using a dimensionless parameter,
\begin{align}
\Omega= \left(\rho_{0} + \rho_{1}\right) \langle \hat{H}_{0,1}^2 \rangle^{1/2},
\end{align}
\emph{i.e.}, the root-mean-square coupling between $J=0$ and $1$ in units of the mean level spacing.
At $\Omega\ll 1$ angular momentum is conserved and the sticking time is set by the $J=0$ density of states,
whereas at $\Omega\gg 1$ angular momentum is scrambled and the sticking time determined by the total density of states.
At intermediate $\Omega$, we calculate the distribution of time delays as described in detail in the Supplement \cite{supp}.
This distribution is then fit with an effective density of states $\rho_\mathrm{eff}$ that is intermediate between these two extremes.
As shown in Fig.~\ref{fig:Omega}, we find the effective density of states increases from $\rho_{J=0}$ at low fields, and approaches the total density of states at high field, as expected.
The transition occurs around $\Omega=1$, where the Stark coupling is comparable to the level spacing.
The electric field at which this occurs in different systems depends strongly on the sticking time, which scales steeply with the masses and the number of degrees of freedom.
For NaK+K this amounts to an electric field of $\sim 10$~kV/cm,
whereas for sticky collisions between typical diatomic polar molecules electric fields in the order of $\sim 10$~V/cm break total angular momentum conservation,
as summarized in Table~\ref{table}.

\begin{table*}
\begin{center}
\caption{ \label{table}
{\bf Summary of numerical results.}
The static limits of the Fourier transform of the dipole and spherical harmonic ACFs, $S_d(0,0)$ and $S_{Y_2}(0,0)$, are listed.
From this we determine the electric field around which the transition to non-conservation of total angular momentum occurs, $E_{\Omega = 1}$,
and find $\Omega_\mathrm{nucl. spin} \ll 1$ indicating nuclear spin is conserved in sticky collisions.
We also list the rates of spontaneous emission and absorption of black-body radiation.
These results are subject to the ambiguity in the precise position of the transition as well as the uncertainty in the available interaction potentials, see Supplement \cite{supp}.
}
\begin{tabular}{lcccccc}
\hline\hline
System    & $S_d(0,0)$ (Debye$^2$ ps)     & $E_{\Omega = 1}$   & $S_{Y_2}$(0,0) (fs)   & $\Omega_\mathrm{nucl. spin}$ & $\Gamma_\mathrm{spont}$ (s$^{-1}$) & $\Gamma_\mathrm{black body}$ (s$^{-1}$) \\
\hline
NaK+K     & 1.2                           & 14.8 kV/cm          &                       &                              &  0.015                             &  0.082     \\
NaK+NaK   & 3.7                           & 48   V/cm          & 28                    &  0.003                       &  0.049                             &  0.26      \\
RbCs+RbCs & 6.1                           & 3.7  V/cm          & 71                    &  0.06                        &  0.0023                            &  0.030     \\
\hline \hline
\end{tabular}
\end{center}
\end{table*}

In the Supplement~\cite{supp} we consider three alternative approaches to determine the transition to $J$ non-conservation.
First, we use a random matrix theory description of the short-range Hamiltonian,
where we observe the level-spacing statistics as a function of the electric field.
With increasing field strength, we observe a transition from Poisson to Wigner-Dyson statistics roughly around $\Omega=1$, consistent with the above analysis.
Second, we employ the framework of Ref.~\cite{christianen:21} to use this random matrix theory description to compute a collisional loss rate.
We set the short-range loss $\Gamma$ such that $\Gamma \rho_{J=0} \ll 1$, which results in a small collisional loss rate.
As we apply an external electric field, the \emph{effective} density of states increases and a transition can be observed in the collisional loss rate as $\Gamma \rho_\mathrm{eff} \approx 1$.
Also this transition occurs roughly around $\Omega = 1$.
Finally, we directly simulated the classical dynamics in an external field.
While $J$ may no longer be conserved during the complex' sticking time, it may be conserved during the short durations we can actually simulate.
Hence, our approach is to record the classical total angular momentum $J(t)$ during the simulation, and fit to $J \propto \sqrt{t}$, consistent with diffusion in total angular momentum.
We then solve for the electric field strength at which $J(t)$ diffuses by one quantum during the sticking time,
which is once again roughly consistent with $\Omega=1$ seen in the other approaches.

The predicted electric field strengths can be tested in experiments, and are consistent with observations so far.
In the RbCs experiment of Ref.~\cite{gregory:20}, which yielded sticking times consistent with angular momentum conservation,
an upper limit to the electric field is 7~V/cm \cite{cornish_private}.
Similarly, in the KRb experiment of Ref.~\cite{liu:20}, a static field of 17~V/cm was present,
which should not break angular momentum conservation given the sticking time is shortened by three orders of magnitude due to chemical reactions.

We emphasize that we have here described the effect of a static electric field on the short-range physics.
This is complimentary to a recent study of the effect of external fields on the long-range physics~\cite{quemener:22}.
The effects on both the long-range and short-range physics can be described in a unified manner~\cite{croft:20,christianen:21}.

\begin{figure}
\centering
\includegraphics[width=0.475\textwidth]{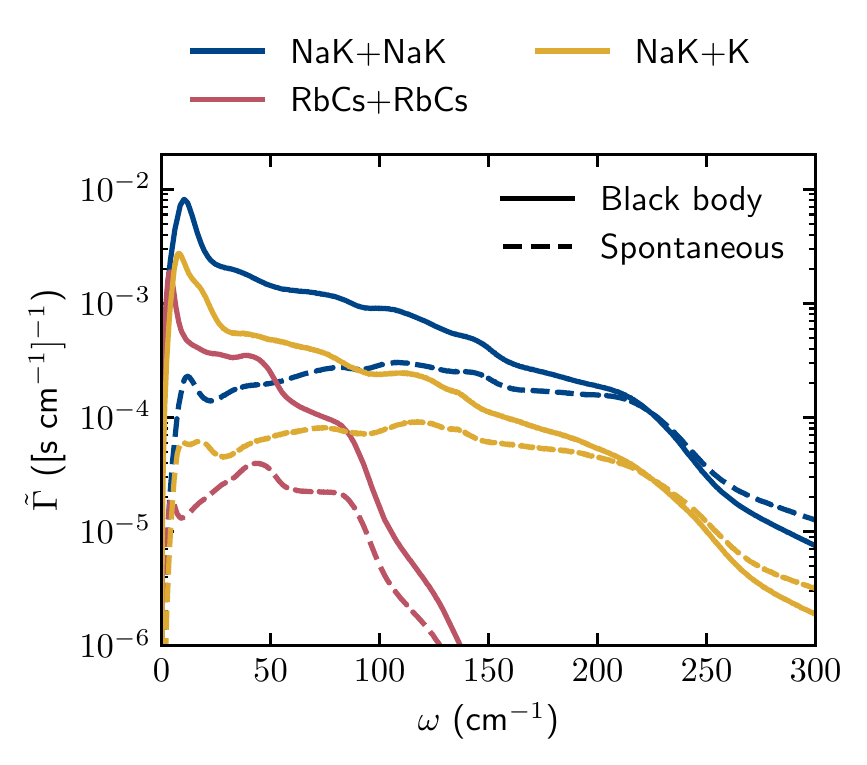}
\caption{
{\bf Spectrum of spontaneous emission and absorption of black-body radiation at 293 K.}
The integrated intensity yields the loss rate,
which is in the order of 0.1~s$^{-1}$ and can essentially be excluded during sticky collisions, see Table~\ref{table}.
}
\label{fig:spectrum}
\end{figure}

Having calculated the dipole autocorrelation function we are also in a position to discuss absorption or emission of far IR radiation by collision complexes.
Figure~\ref{fig:spectrum} shows the absorption spectrum of black-body radiation at 293~K, and the spectrum of spontaneous emission.
The rates of of both processes are on the order of 0.1~s$^{-1}$, see Table~\ref{table}, and can be ruled out on the time scale of sticky collisions.

Finally, we apply the formalism developed here to the conservation of the nuclear spin state during a collision.
For simplicity, we limit the discussion to the strongest hyperfine interaction.
For RbCs this is the quadrupole coupling of the Rb nuclear spin $i=3/2$ to the electric field gradient.
Apart from constants, this takes the form $[\ [\ \hat{i} \otimes \hat{i}\ ]^{(2)} \otimes Y^{(2)}(\hat{r})\ ]^{(0)}$,
where $\hat{i}$ is a nuclear spin operator, $[\hat{A} \otimes \hat{B} ]^{(k)}$ is a rank-$k$ tensor product, and $Y^{(2)}(\hat{r})$ is a tensor of spherical harmonics depending on the polar angles of the molecular axis, $\hat{r}$.
As with the dipole moment before, the dynamics of the collision complex rapidly reorients the molecular axis, resulting in a fluctuating coupling for the nuclear spin.
The fluctuations are characterized by the ACF of the spherical harmonics, which determines the coupling from an initial spin state, say $m_i=3/2$, to $m_i'=1/2$ and $-1/2$.
The $|\Delta m_i| \le 2$ selection rule results from the second-rank coupling, which also changes the mechanical angular momentum from $J=0$ to $J'=2$.
Numerical results for the relevant ACFs are shown in Fig.~\ref{fig:ACF}.
From our simulations shown in Fig.~\ref{fig:Omega} we can then immediately conclude nuclear spin is conserved for $\Omega \ll 1$.
For RbCs, using hyperfine coupling constants form the literature and the simulated autocorrelation function of the spherical harmonics, we find $\Omega \approx 0.06$ such that nuclear spin remains conserved.
A similar analysis accounting for the K nuclear quadrupole coupling in NaK+NaK complexes yields $\Omega = 0.003$, see Table~\ref{table}.
This is in qualitative agreement with experimental sticking times that match theory assuming spin conservation \cite{gregory:20,liu:20},
and consistent with spin-conservation in KRb+KRb collisions \cite{hu:21},
although recent results in RbCs collisions suggest the sticking time might be hyperfine state dependent \cite{gregory:21}.
The approach developed here opens the door to subsequent studies accounting for all hyperfine couplings, their interaction-induced variations \cite{jachymski:21},
and to explore the case of non-zero electronic spin where the finestructure couplings may be orders of magnitude larger.

In conclusion, we have developed classical simulations of sticky collisions between ultracold molecules.
The dynamics is consistent with RRKM theory and the sticking times predicted previously.
We show how the non-conservation of nearly good quantum numbers (such as total angular momentum) can be calculated using the autocorrelation function of a perturbation (such as coupling to an $E$-field).
Static electric fields as small as $10$~V/cm are found to lead to non-conservation of total angular momentum.
The same dipole ACF can be used to study loss of collision complexes by spontaneous emission or absorption of black-body radiation, which we conclude is slow.
In addition, we show how the same method may be applied to the non-conservation of nuclear spin,
suggesting the nuclear spin is conserved in sticky collisions, tentatively in agreement with observations.
The framework presented here creates new possibilities to quantitatively study loss processes in ultracold collision complexes.
By understanding the sticking times and loss processes quantitatively, we can hope to eliminate collisional loss of ultracold molecules which will aid the creation of long-lived molecular quantum gases,
collisional cooling,
and support Feshbach resonances that enable control of short-range interactions.
The method employed here is quite general may also be used to efficiently simulate rare events in other areas of physics.

\section*{Acknowledgements}

We thank Simon Cornish for useful discussions.
Funding for this work was provided by the Dutch Research Council (NWO) (Vrije Programma 680.92.18.05).

\bibliography{vanderwaals,extras,papersHerma}

\newpage
\onecolumngrid
\begin{center}
{\bf \large Supplemental Material}
\end{center}

\section{Primer on notation}

Throughout this manuscript and Supplementary Material we encounter multiple timescales with distinct physical meanings.
Examples are the sticking time of an ultracold collision complex, $\tau_\mathrm{stick}$,
and the time required before the molecules cross a dividing surface $R=R_S$, $\tau_S$, for some specific value of $R_S$.
We denote these various timescales by the symbol $t$ or $\tau$, and where applicable indicate the physical meaning by subscripts, such as $\tau_\mathrm{stick}$, $\tau_S$, and so on.

In this work, we also compare Rice-Ramsperger-Kassel-Marcus (RRKM) theory to classical trajectory simulations.
We use superscripts ``RRKM'' and ``traj'' in order to distinguish the same physical quantity computed using these two methods.
For example, $\tau_\mathrm{stick}^\mathrm{RRKM}$ and $\tau_\mathrm{stick}^\mathrm{traj}$ denote the same physical quantity, the sticking time of an ultracold collision complex, but computed using RRKM theory and trajectory simulations, respectively.

\section{Sticking times from phase-space integrals \label{sec:phasespace_lifetime}}

Mayle et al.~\cite{mayle:12,mayle:13} have proposed that ultracold collision complexes are described by Rice-Ramsperger-Kassel-Marcus (RRKM) theory or Transition State Theory \cite{Rice:1927,Kassel:1928,eyring:35,marcus:52}.
The central result is that the time spent in a phase-space region is given by
\begin{align}
\tau_S^\mathrm{RRKM} = \frac{2\pi \hbar \rho_S}{N_S},
\label{eq:rrkm}
\end{align}
where $\rho_S$ is the density of states in that phase-space region, and $N_S$ is the number of states at the boundary \cite{miller:93}.
This boundary will be referred to as the dividing surface.
The physical picture is that the collision complexes ergodically explore phase-space until they find and cross the dividing surface and dissociate.
The time that this process takes, as described by Eq.~\eqref{eq:rrkm}, increases with the density of states that the complex explores chaotically,
and decreases with the number of states at the dividing surface, as this increases the chance this ``transition state'' is encountered.

Assuming chaotic dynamics, the ``sticking time'' of the collision complex, Eq.~\eqref{eq:rrkm}, is determined only by the density of states inside the dividing surface, $\rho_S$, and the number of states at the dividing surface, $N_S$.
Both $\rho_S$ and $N_S$ are calculated using phase-space integrals \cite{miller:93,christianen:19b}.
For $\rho$ we use the formulas from Christianen \textit{et al.}~\cite{christianen:19b}.
For collisions between a diatom AB and an atom B we have
\begin{equation}
    \rho=g_{NJp}\int_0^\pi\!\!\int_0^\infty\!\!\int_0^\infty\!\frac{2(2\pi)^{3/2}\sqrt{\mu_{\text{A},\text{B}}\mu_{\text{AB},\text{B}}I_\text{3at}}}{\sqrt{\pi}\prod_iN_i!(2\pi\hbar)^3}\sqrt{E-V(r,R,\theta)} \heaviside[V(r,R,\theta)-E]dr\,dR\,d\theta, 
\end{equation}
where
\begin{equation}
    I_\text{3at}=[\mu_{\text{A},\text{B}}^{-1}r^{-2}+\mu_{\text{AB},\text{B}}^{-1}R^{-2}]^{-1},
\end{equation}
$\heaviside(\cdot)$ is the Heaviside step function, $\mu_{A,B}$ is the reduced mass of particles $A$ and $B$, $g_{NJp}$ is a factor that accounts for parity\cite{christianen:19b}, $N_i$ is the number of identical particles of type $i$, $V(r,R,\theta)$ is the potential energy surface (PES), ($r$, $R$, $\theta$) are Jacobi coordinates, and $E$ is the total energy of the system. We set $E = \lim_{R\rightarrow \infty}V(r_e,R,\theta)$, with $r_e$ is the equilibrium distance of diatom $\text{AB}$, \emph{i.e.}, we set the collision energy to zero.
Throughout this work we use diatomics-in-molecules (DIM) PESs, described in detail in Sec.~\ref{sec:DIM}.
For collisions between diatoms $\text{AB}$ and $\text{CD}$ we have
\begin{equation}
    \rho=\int\frac{g_{NJp}(m_\text{A}m_\text{B}m_\text{C}m_\text{D})^3R^4r_1^4r_2^4\sin^2{(\theta_1)}\sin^2{(\theta_2)}}{\pi^3 2^{7}\prod_iN_i!(m_\text{A}+m_\text{B}+m_\text{C}+m_\text{D})^3\det{I_\text{4at}}\sqrt{\det{A}}}[E-V(\mathbf{q})]^2 \heaviside[V(q)-E]d\mathbf{q}, 
\label{eq:rho4}
\end{equation}
where $\mathbf{q}$ is a set of diatom-diatom Jacobi coordinates.
The integration bounds are $\theta_1$, $\theta_2 \in [0,\pi ]$, $\phi \in [0,2\pi ]$, and $R$, $r_1$, $r_2 \in (0, \infty )$.
Furthermore, $m_{\text{A}/\text{B}/\text{C}/\text{D}}$ is the mass of particle $\text{A}/\text{B}/\text{C}/\text{D}$, and both $I_\text{4at}$ and $\sqrt{\det{A}}$ are defined in Ref.~\cite{christianen:19b}.
In the case of $\mathrm{AB}+\mathrm{AB}$ collisions we have $g_{NJp}=\frac{1}{2}$ and $\prod_iN_i!=4$. 

The number of states at the dividing surface $R=R_S$ is given for $\mathrm{AB}+\mathrm{B}$ by
\begin{equation}\label{eq:N_3at}
    N_S=M_\text{exit}g_{NJp}\int_0^\pi\!\!\int_0^\infty\!\frac{2\pi\sqrt{\mu_{\text{A},\text{B}}I_\text{3at}}}{(2\pi\hbar)^2\prod_iN_i!}[E-V(r,R,\theta)] \heaviside[V(r,R,\theta)-E]dr\,d\theta ,
\end{equation}
and for $\mathrm{AB}+\mathrm{AB}$ by
\begin{equation}\label{eq:N_4at}
    N_S=M_\text{exit}g_{NJp}\int\frac{(m_\text{A}m_\text{B})^6 R^4r_1^4r_2^4\sin^2{(\theta_1)}\sin^2{(\theta_2)}[E-V(r,R,\theta)]^{5/2} \heaviside[V(\mathbf{\tilde{q}})-E]}{15\pi^3 2^{5}(m_\text{A}+m_\text{B})^3\prod_iN_i!\mu_{\text{AB},\text{AB}}\det{I_\text{4at}}\sqrt{\det{A}}}d\mathbf{\tilde{q}}.
\end{equation}
Here $\mathbf{\tilde{q}}$ is the set of $\mathrm{AB}+\mathrm{CD}$ Jacobi coordinates excluding the $R$-coordinate.
The factor $M_\text{exit}=2$ accounts for dissociation in the two equivalent arrangements $\mathrm{AB}+\mathrm{B}'$ and $\mathrm{AB}'+\mathrm{B}$ for the atom-diatom case and $\mathrm{AB}+\mathrm{A}'\mathrm{B}'$ and $\mathrm{A}\mathrm{B}'+\mathrm{A}'\mathrm{B}$ for diatom-diatom.
It is assumed that chemical reactions are energetically not allowed and that the corresponding arrangements do not contribute to the number of exit channels.
We use the same integration bounds as for the density of states. 
The phase-space integral for $N_S$ can be derived using the same steps as those taken by Christianen \textit{et al.}~\cite{christianen:19b},
but restricting the coordinates to the dividing surface $R=R_S$.
This amounts to fixing $R=R_S$, omitting its conjugate momentum, and multiplying the result by $2\pi\hbar$ \cite{miller:93}.
The factors $g_{NJp}$ and $\prod_iN_i!$ account for parity and identical particle symmetry.
Since these factors are applied to both $\rho$ and $N_S$, they have no effect on the sticking time of Eq.~\eqref{eq:rrkm}.

\begin{figure}
    \centering
    \includegraphics[width=0.7\textwidth]{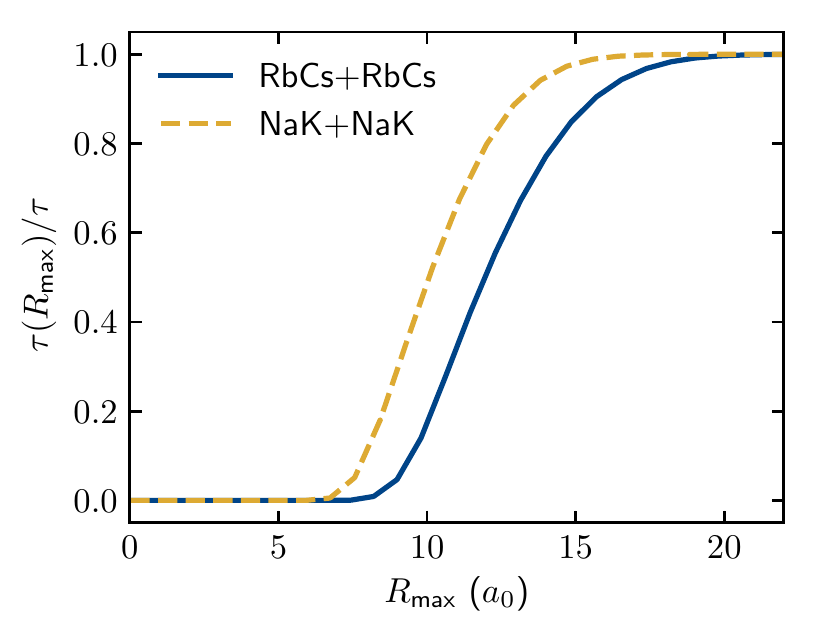}
    \caption{
Contribution of $R \le R_\mathrm{max}$ to the density of states Eq.~\eqref{eq:rho4}, normalized to the total density of states.
This can be interpreted as the time the complex spends in the region $R\le R_\mathrm{max}$, $\tau(R_\mathrm{max})$, normalized to the sticking time.
    \label{fig:partial_tau}}
\end{figure}

In order to get an idea of the structure of these chaotic collision complexes, the Jacobi coordinate $R$ can be restricted to $R \le R_\mathrm{max}$ to compute the contribution to the density of states in Eq.~\eqref{eq:rho4}.
Figure~\ref{fig:partial_tau} shows this restricted density of states normalized to the total density of states, or equivalently the restricted sticking time normalized to the total sticking time,
as a function of $R_\mathrm{max}$ for NaK+NaK and RbCs+RbCs complexes.
This illustrates that the sticky collision complexes spend most of their time at distances shorter than 15 to 20~$a_0$.
This suggests a clear separation of length scales of the short-range physics occurring at $R\le 20~a_0$,
and the long-range physics that takes place on the van der Waals length scale that is typically hundreds of $a_0$ for molecule-molecule collisions \cite{gao:98}.
Hence, one can imagine placing the dividing surface at some $R_S$ between these two length scales,
which unambiguously defines the density of states in Eq.~\eqref{eq:rrkm}.

The number of states at the dividing surface, $N_S$, is not independent of $R_S$, however, since
\begin{equation}\label{eq:N_lim_cl}
    \lim_{R_S\rightarrow\infty}N_{S}=0
\end{equation}
at $E=0$. That is, the classical sticking time at zero energy diverges.
This can be contrasted with the quantum mechanical situation where
\begin{equation}
    \lim_{R_S\rightarrow\infty}N_{S}^\text{QM} = 1,
\end{equation}
since in ultracold systems the lowest $s$-wave channel would remain energetically accessible at large distances $R$~\cite{mayle:12,mayle:13,croft:14}.
Hence, a naive application of the classical RRKM formula to ultracold collisions would lead to unphysical results.
The physical sticking times are therefore obtained from Eq.~\eqref{eq:rrkm} with $N_S=1$.

\section{Sticking times from classical trajectory simulations \label{sec:trajectory_lifetime}}

\subsection{Theory}

We initialize the trajectories with both molecules at random orientations and with zero rotational and vibrational energy.
The molecules are initially far apart, moving towards each other with kinetic energy chosen such that the total energy is equal to the lowest dissociation limit.
We then simulate the trajectory until the system crosses a dividing surface to $R>R_S$.
The time that this takes, $\tau_S^\mathrm{traj}$, is averaged over many trajectories.
See Sec.~\ref{sec:trajectory_lifetime:compdet} for details.

As discussed in the previous section, in a purely classical calculation the RRKM sticking time diverges as the dividing surface $R=R_S$ is moved outwards,
because the number of states at the dividing surface at zero energy vanishes as $R_S \rightarrow \infty$.
The sticking time from classical trajectory simulations diverges for the same reason.
However, one can still compare the finite time $\tau_S$ it takes to cross a fictitious dividing surface $R=R_S$ for finite $R_S$ as predicted by RRKM theory, $\tau_S^\mathrm{RRKM}$, and trajectory simulations, $\tau_S^\mathrm{traj}$, respectively.

In reality, the exit channels are quantized and the $s$-wave channel remains accessible at ultracold temperatures such that $N_S^\mathrm{QM}$ approaches 1 rather 0, as $R_S\rightarrow \infty$.
Hence, at the RRKM level, the physical sticking time can be calculated by setting $N_S=1$ in Eq.~\eqref{eq:rrkm},
which yields
\begin{align}
\tau_\mathrm{stick}^\mathrm{RRKM} &= \lim_{R_S\rightarrow \infty} 2\pi\hbar \rho_S \nonumber \\
&= \lim_{R_S\rightarrow \infty} N_S \tau_S^\mathrm{RRKM}.
\end{align}
We assume $\tau_S^\mathrm{traj}$ has the same $1/N_S$ scaling which enables estimating the physical sticking time as
\begin{align}
\tau_\mathrm{stick}^\mathrm{traj} = N_S \tau_S^\mathrm{traj}.
\label{eq:rrkmtraj}
\end{align}
We then compare the sticking times estimated from classical trajectory simulations for various finite values of $R_S$.
This comparison is shown in Fig.~1(b) of the main text, which shows that different choices of $R_S$ result in widely different values for $N_S$, but lead to a constant estimate of the sticking time, $\tau_\mathrm{stick}^\mathrm{traj}$, that is consistent with RRKM theory.

The approach taken here has the additional advantage that we can choose $R_S$ between 15 and 24~$a_0$, such that $N_S \gg 1$.
This implies the classical approximation is reasonable,
and it leads to the computational advantage that the typical time before the trajectory crosses the dividing surface, $\tau_S^\mathrm{traj}$, is shorter.
As a result, our approach enables the study of sticky collisions by simulating classical trajectories for times that are much shorter than the sticking time.

\subsubsection{Ergodic behavior}

The classical trajectory calculations probe whether ultracold collision complexes are ergodic, which is assumed in RRKM theory.
However, our computational approach might exaggerate the effects of nonergodic behavior.
Initially, all kinetic energy is in the relative motion of the two molecules, not their rotation or vibration.
If the dynamics is chaotic, this energy will redistribute among the different degrees of freedom.
However, in our simulations we use dividing surfaces at relatively short $R_S$ such that $N_S\gg 1$,
and it is possible that the simulated trajectories do not become ergodic before the first crossing,
even if the trajectory would ``thermalize'' and become ergodic before actually dissociating.
Hence, during a typical collision the chosen dividing surface could be crossed many times before dissociation, and using the first crossing might lead to an underestimation of the sticking time.
We investigate this by first running the trajectories for a ``thermalization time'', $t_\mathrm{therm}$, before we start recording the time between the first and second crossings to $R>R_S$.

We then investigate the effect of nonergodic behavior on the sticking time by determining the fraction of the trajectories that lead to dissociation during the thermalization time.
We consider the complex dissociated if $R>24.79~a_0$ for NaK+NaK and $R>32.25~a_0$ for RbCs+RbCs.
At these intermolecular distances, $N_S \approx 1$.
Trajectories that dissociate are not taken into account when calculating the sticking time.
The fraction of dissociated trajectories is considered a measure of nonergodic behavior.

For NaK+K the sticking time is much shorter, so we can simulate the full physical sticking time by choosing $R_S=22.52~a_0$ such that $N_{S}\approx 1$.
We then probe the influence of nonergodic behavior by comparing the sticking time calculated with and without thermalization, see Sec.~\ref{sec:trajectory_lifetime:res}.

\subsection{Computational details \label{sec:trajectory_lifetime:compdet}}

We ran classical trajectory calculations for three systems; $^{23}$Na$^{39}$K+NaK, $^{87}$Rb$^{133}$Cs+RbCs, and NaK+K.
We propagated the trajectories using the fourth-order symplectic PEFRL algorithm \cite{omelyan:02}. In the sticking time calculations (except the calculation where we added the zero-point energy (ZPE) to the collision energy) we used time steps of $1.2~10^{-6}$~ns, $1.2~10^{-6}$~ns, and $1.5~10^{-8}$~ns for NaK+NaK, RbCs+RbCs, and NaK+K, respectively.
In all other calculations we used time steps of $4.8~10^{-6}$~ns, $4.8~10^{-6}$~ns, and $2.5~10^{-8}$~ns for NaK+NaK, RbCs+RbCs, and NaK+K, respectively.
We use a diatomics-in-molecules (DIM) potential that is described in detail in Sec.~\ref{sec:DIM}.
This describes the interactions as spin-dependent pairwise interactions.
The pair potentials are assumed Morse potentials with the well depths, zero-point energies (ZPE), and equilibrium distances ($r_e$) based on Refs.~\cite{gerdes:08,pashov:07,knoop:11,li:85,docenko:11,krauss:90,amiot:02,xie:09,tomza:13} given in Table \ref{tab:DIM_parameters}.
Forces were calculated numerically using central finite differences with distance $1\cdot 10^{-6}$~$a_0$.
We describe the diatom-diatom systems using ($R,r_1,r_2,\theta_1,\theta_2,\phi$) Jacobi coordinates and the atom-diatom system using ($R,r,\theta$) Jacobi coordinates.
We initialized the four-atom systems by setting $R=\mathrm{27}~a_0$, $r_1=r_2=r_e$, and we sampled $\cos(\theta_1)$, $\cos(\theta_2)$, and $\phi$ uniformly randomly.
We initialized the three-atom system with $R=\mathrm{27}~a_0$, $r=r_e$, and $\cos(\theta)$ was sampled uniformly randomly.
All conjugate momenta were initialized at zero, except for movement in the $R$ direction.
The kinetic energy in the $R$ coordinate was chosen such that the total energy of the system was zero.
This also ensures that initially the total angular momentum was zero. 

Sticking times were calculated by first running the trajectories for the duration of the thermalization time, $t_\text{therm}$.
After this we wait until the dividing surface is crossed to $R>R_S$.
We then set the clock to zero.
Next, we let the trajectories run until the dividing surface is crossed to $R>R_S$ a second time, and record the elapsed time.
The elapsed times of many different trajectories are averaged to obtain $\tau_S^\mathrm{traj}$, the mean time for dissociation through the dividing surface $R=R_S$ in a trajectory calculation.

\begin{table}[]
    \centering
    \caption{Well depths ($D_e$), zero-point energies (ZPE), and equilibrium distances ($r_e$) from the literature that were used to determine the Morse potentials used in the DIM potentials.}
    \begin{tabular}{llSSSl}
    \hline
    \hline
        Diatom &Term symbol & \multicolumn{1}{c}{$D_e$}&\multicolumn{1}{c}{ZPE}&\multicolumn{1}{c}{$r_e$}&Reference\\
        & & \multicolumn{1}{c}{(cm$^{-1}$)}&\multicolumn{1}{c}{(cm$^{-1}$)}&\multicolumn{1}{c}{(\r{A})}&\\
        \hline
        NaK&X$^1\Sigma^+$&5274&61.9&3.50&\cite{gerdes:08}\\
        NaK&a$^3\Sigma^+$&208&11.3&5.45&\cite{gerdes:08}\\
        K$_2$&X$^1\Sigma^+_g$&4451&46.1&3.92&\cite{pashov:07}\\
        K$_2$&a$^3\Sigma^+_u$&255&10.5&5.73&\cite{pashov:07}\\
        Na$_2$&X$^1\Sigma^+_g$&6022&79.4&3.08&\cite{knoop:11}\\
        Na$_2$&a$^3\Sigma^+_u$&173&12.2&5.14&\cite{li:85,knoop:11}\\
        \hline
        RbCs&X$^1\Sigma^+$&3836&24.8&4.43&\cite{docenko:11}\\
        RbCs&a$^3\Sigma^+$&259&6.2&6.22&\cite{docenko:11}\\
        Cs$_2$&X$^1\Sigma^+_g$&3650&21.0&4.63&\cite{krauss:90,amiot:02}\\
        Cs$_2$&a$^3\Sigma^+_u$&279&5.8&6.24&\cite{xie:09}\\
        Rb$_2$&X$^1\Sigma^+_g$&3912&28.1&4.23&\cite{tomza:13}\\
        Rb$_2$&a$^3\Sigma^+_u$&250&6.8&6.06&\cite{tomza:13}\\
        \hline
        \hline
    \end{tabular}

    \label{tab:DIM_parameters}
\end{table}

\subsection{Results \label{sec:trajectory_lifetime:res}}

We validated the trajectory method by calculating the sticking time as a function of the position of the dividing surface $R=R_S$.
We then compared the sticking time obtained for different dividing surfaces to one another.
We also compared to the RRKM sticking time, $\tau_\mathrm{stick}^\mathrm{RRKM}$, from Eq.~\eqref{eq:rrkm} with $\rho$ obtained from phase-space integrals,
\emph{i.e.}, a calculation that is independent of the trajectory simulations.
To this end, we recorded the time, $\tau_S^\mathrm{traj}$, that a trajectory takes to cross the dividing surface to $R>R_S$.
We then estimate the sticking time using Eq.~\eqref{eq:rrkmtraj}, which essentially corrects for the number of states at the dividing surface $N_S$,
if this differs from the single quantized channel that is accessible in ultracold collisions.
In each sticking time calculation we ran 1000 trajectories as described in Sec.~\ref{sec:trajectory_lifetime:compdet}.
We ran these trajectories until their first crossing to $R>R_S$ after the thermalization time and used the average of the elapsed time since the thermalization time, $\tau_S^\mathrm{traj}$, to calculate the sticking time using Eq.~\eqref{eq:rrkmtraj}.
The results are listed in Table \ref{tab:classical_lifetime}.

We found that the sticking times estimated using classical trajectories agree well with the sticking times from RRKM based on phase-space integrals.
This is illustrated in Fig.~1(b) of the main text, where we plot both $\tau_\mathrm{stick}^\mathrm{traj}=N_{S} \tau_S^\mathrm{traj}$ and $N_{S}$ for several values of $R_{S}$ for NaK+NaK.
This shows that the trajectory method gives reasonable sticking times $\tau_\mathrm{stick}^\mathrm{traj}$ for dividing surfaces over a range of positions $R_{S}$,
even when the number of channels at the dividing surface, $N_{S}$, varies by orders of magnitude.
The typical time a simulated trajectory takes to cross to $R>R_S$ also varies by orders of magnitude, but the estimated sticking time $\tau_\mathrm{stick}^\mathrm{traj}$ is essentially independent of $R_S$ and in agreement with RRKM theory.
This validates the applicability of RRKM, which assumes ergodic dynamics.
We use our classical trajectories to further investigate the effects of nonergodic behavior.
For NaK+NaK and RbCs+RbCs collisions, the fraction of the trajectories that dissociates during thermalization, $f_\text{dis}^\text{therm}$, is less than 2\%, validating the ergodic assumption of RRKM.
Similarly for NaK+K collisions we find that the sticking times from classical trajectories with and without thermalizing the system are similar, again indicating that the ergodic assumption is valid.

For NaK+NaK, we also ran $1000$ trajectories without thermalization and with additional kinetic energy, initially in the $R$-coordinate, such that the total energy of the system is equal to the combined ZPE of the two NaK diatoms of $123.8$~cm$^{-1}$.
Note this makes chemical reactions NaK$+$NaK$\rightarrow$Na$_2+$K$_2$ allowed as this reaction is now exothermic by 49~cm$^{-1}$.
We used $R_{S}=24.79~a_0$.
For this dividing surface, at zero energy, we would have $N_S\approx 1$.
However, when we calculate $N_S$ including the additional ZPE energy, we obtain $N_S^\text{ZPE}=807$.
Since the Na$_2$+K$_2$ reaction is now energetically accessible we removed the trajectories when the $R$ Jacobi coordinate associated with the Na$_2$+K$_2$ configuration is larger than $100$ $a_0$.
We removed $18$ trajectories, these trajectories are not taken into account when calculating the sticking time.
We also removed one additional trajectory after it ran for an unrealistically long time.
We find the trajectories escape after $\tau_S^\mathrm{traj}=25.6$~ns.
This is much shorter than the ultracold sticking time.
However, when we account for the higher $N_S^\text{ZPE}$ due to the additional ZPE energy, we get the estimate of the sticking time $\tau^\mathrm{traj}_\text{stick} = N_S^\text{ZPE} \tau_S^\mathrm{traj} = 20.7$~$\mu$s,
which is close to the RRKM sticking time.
Thus, adding additional kinetic energy drastically shortened the sticking time, but only by the trivial effect that additional exit channels become energetically accessible.
When we compensated for this phenomenon, the sticking time is close to the original estimate.

\begin{table}[]
    \centering
    \caption{For several systems and dividing surfaces we list the position of the dividing surface, $R_{S}$, the mean time (and 1-$\sigma$ statistical uncertainty) between crossings of the dividing surface, $\tau_S^\mathrm{traj}$, the number of states at the dividing surface, $N_{S}$, the sticking time estimated from trajectory simulations using $\tau_\text{stick}^\text{traj}=\tau_SN_{S}$, the sticking time calculated using RRKM from phase-space integrals, $\tau^\text{RRKM}_\text{stick}$, the thermalization time, $t_\text{therm}$, and the fraction of the trajectories that dissociated during thermalization, $f_\text{dis}^\text{therm}$. Sticking times from trajectory simulations and RRKM theory are in good agreement. Note that the sticking times in this table differ substantially from previous calculations~\cite{christianen:19b} and experiment~\cite{gregory:20}, reflecting the accuracy of our DIM interaction potentials.}
    \begin{tabular}{lSSSSSSS}
    \hline
    \hline
         System&\multicolumn{1}{c}{$R_{S}$}&\multicolumn{1}{c}{$\tau_S^\mathrm{traj}$}&\multicolumn{1}{c}{$N_{S}$}&\multicolumn{1}{c}{$\tau^\text{traj}_\text{stick}$}&\multicolumn{1}{c}{$\tau^\text{RRKM}_\text{stick}$}&\multicolumn{1}{c}{$t_\text{therm}$}&$f_\text{dis}^\text{therm}$\\
         &\multicolumn{1}{c}{($a_0$)}&\multicolumn{1}{c}{(ns)}&&\multicolumn{1}{c}{($\mu$s)}&\multicolumn{1}{c}{($\mu$s)}&\multicolumn{1}{c}{(ns)}&\\
         \hline
         NaK+NaK&20&22.0&802&17.7    (6)&18.5&20&0.006\\
         NaK+NaK&19&5.8&3166&18.2    (6)&18.5&20&0.011\\
         NaK+NaK&18&1.5&12206&18.3   (6)&18.5&20&0.009\\
         NaK+NaK&17&0.44&45086&19.9  (6)&18.5&20&0.005\\
         NaK+NaK&16&0.12&153232&18.7 (6)&18.5&20&0.007\\
         NaK+NaK&15&0.041&448667&18.4(6)&18.5&20&0.006\\
         \hline
         RbCs+RbCs&24&52.0&34749&1806 (57) &1869&4&0.011\\
         \hline
         &\multicolumn{1}{c}{($a_0$)}&\multicolumn{1}{c}{(ns)}&&\multicolumn{1}{c}{(ns)}&\multicolumn{1}{c}{(ns)}&\multicolumn{1}{c}{(ns)}&\\
         \hline
         NaK+K&22.52&0.65&1.0&0.65 (2) &0.62&2&\multicolumn{1}{c}{-}\\
         NaK+K&22.52&0.69&1.0&0.69(2) &0.62&0&\multicolumn{1}{c}{-}\\
    \hline
    \hline
    \end{tabular}
    \label{tab:classical_lifetime}
\end{table}

\section{Classical trajectories in electric fields}

\begin{figure}
    \centering
    \includegraphics[width=0.7\textwidth]{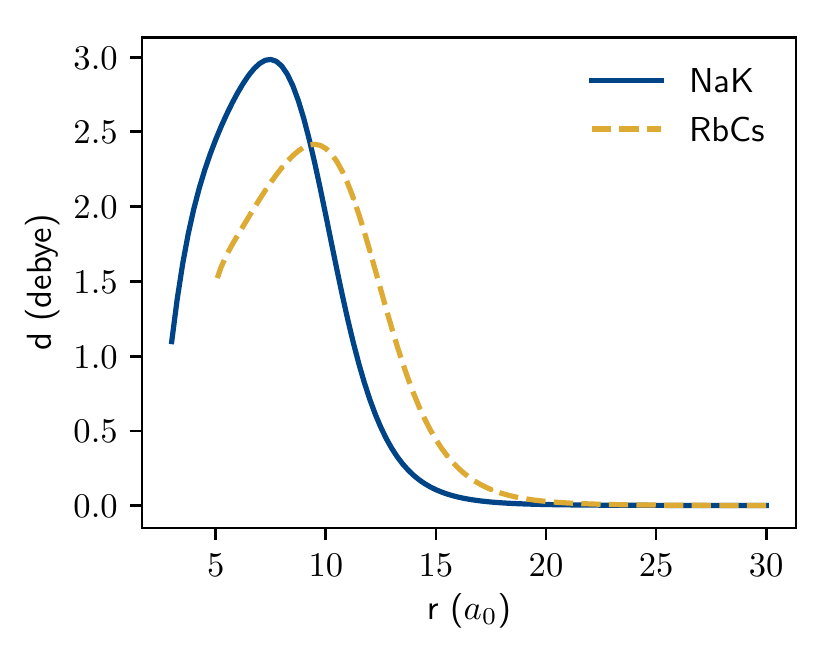}
    \caption{Dipole moments of NaK and RbCs as a function of the internuclear distance, $r$. }
    \label{fig:dipole_diatom}
\end{figure}

When simulating collision complexes in electric fields, we added the interaction between the dipole moment of the complex, $\bm{d}_\text{tot}$ and the electric field, $\bm{E}_\text{elec}$, to the Hamiltonian
\begin{equation}
    V=-\bm{E}_\text{elec}\cdot \bm{d}_\text{tot}.
    \label{eq:V_electric_field}
\end{equation}
We estimated the total dipole moment as the vector sum of the dipole moments of the constituent diatoms
\begin{equation}
    \bm{d}_{ij}=\sum_{i\in \text{atoms}} \sum_{j\neq i} \mathbf{\hat{r}}_{ij}d(i,j,r_{ij}),
\end{equation}
where $d(i,j,r_{ij})$ is the dipole moment of the diatom consisting of atoms $i$ and $j$ at internuclear distance $r_{ij}$.
The unit vector $\hat{\mathbf{r}}_{ij}$ points from atom $i$ to atom $j$.
The dipole moment of the diatom was calculated on a grid from $3$ to $30~a_0$ for NaK and $5$ to $30~a_0$ for RbCs, in steps of $0.25$ $a_0$.
The dipole moment is assumed zero for $r>30~a_0$,
and constant for $r$ shorter than the first grid point.
Between grid points the dipole moment was calculated using linear interpolation.
The dipole moment at the grid points was calculated with the MOLPRO 2015.1 program \cite{werner:11,molpro}, using full configuration interaction (FCI) in the valence electrons, with large-core effective-core potentials (ECPs) and core-polarization potentials (CPPs) from Refs.~\cite{fuentealba:82,vonszentpaly:82,fuentealba:83}. 
Hence, only a single valence electron per atom is treated explicitly.
We used basis sets from Ref.~\cite{christianen:19}.
The resulting dipole moments are displayed in Fig.~\ref{fig:dipole_diatom}.

When no electric field is present the total angular momentum of the collision complex is conserved, but when an electric field is introduced it can change. We simulate this process using classical trajectories. We use binning \cite{christianen:19b} to assign the classical angular momentum a discrete quantum number.
The bin for quantum number $J_\mathrm{QM}$ ranges from $B(J_\mathrm{QM})$ to $B(J_\mathrm{QM}+1)$ where
\begin{equation}
    B(J)= \hbar\sqrt[3]{\frac{1}{4}J(2J-1)(2J+1)}.
\end{equation}
One can then imagine simulating non-conservation of angular momentum by (1) randomly initializing trajectories with $J=0$ as before,
(2) running these trajectories for the duration of their sticking time,
and (3) after this time, binning the angular momentum and recording the fraction of trajectories assigned $J_\mathrm{QM}>0$.

Unfortunately running trajectories for this long is computationally too expensive for systems with long sticking times.
Instead, we ran trajectories starting with $J=0$ for a shorter time and fitted the angular momentum at time $t$ to
\begin{equation}
    \mathbb{E}[J(t)]=\alpha E_\mathrm{elec} \sqrt{\frac{t\pi}{2}}, 
    \label{eq:expectation_total_angular_momentum}
\end{equation}
where $\alpha$ is a constant to be fitted.
This enabled extrapolating these results to later times.
Equation~\eqref{eq:expectation_total_angular_momentum} can be justified as follows.
We assume the dipole moment of the collision complex randomly changes orientation on a short time scale.
The total angular momentum along the field direction, $z$, is conserved,
while the $x$ and $y$ components are described by a two-dimensional random walk
\begin{equation}\label{eq:Wiener}
    J_{x/y}(t)\sim N(0,\alpha^2E^2t), 
\end{equation}
where $N(\mu,\sigma^2)$ is the normal distribution (with mean $\mu$ and variance $\sigma^2$) and $\alpha$ is a constant that is fitted to our classical trajectories. 
This implies Eq.~(\ref{eq:expectation_total_angular_momentum}), which can then be used to fit the ``diffusion constant'' $\alpha$. 

We can then assume the total angular momentum states with $J_\text{QM} > 0$ become relevant if the expectation value of the total angular momentum after the sticking time is binned to a non-zero quantum number.
This gives the critical electric field strength
\begin{equation}
    E_\text{elec}^\text{traj}=\frac{\sqrt{2}\sqrt[3]{\frac{3}{4}}}{\alpha\sqrt{\tau_{J=0}\pi}}\hbar. 
\end{equation}

\begin{figure}
    \centering
    \includegraphics[width=0.7\textwidth]{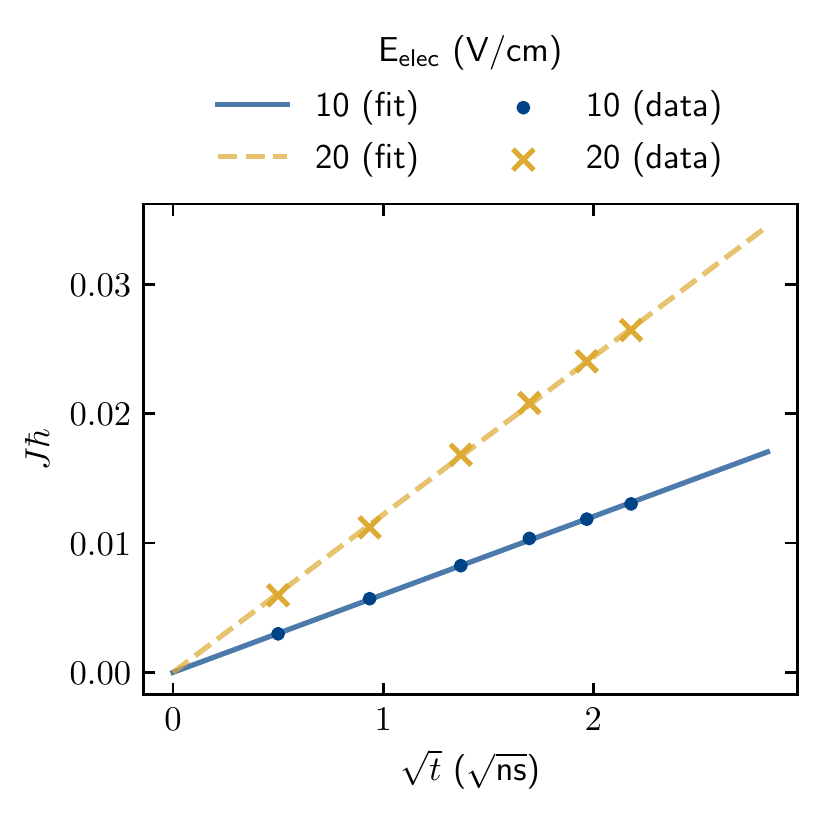}
    \caption{Total angular momentum $J$ of NaK+NaK in electric fields of $10$ or $20$ V/cm as a function of the square root of the run time, averaged over 1\,000 trajectories. 
The data is fit accurately by the $\sqrt{t}$ dependence of Eq.~\eqref{eq:expectation_total_angular_momentum}.
    \label{fig:classical_J}}
\end{figure}

We ran an additional 1000 trajectories for 5~ns for electric field strengths  $E_\text{elec}$=$2.5$, $5$, $10$, and $20$ V/cm.
We recorded the total angular momentum every $0.25$~ns.
A fit of these data to Eq.~\eqref{eq:expectation_total_angular_momentum} is shown in Fig.~\ref{fig:classical_J}.
The quality of this fit empirically justifies Eq.~\eqref{eq:expectation_total_angular_momentum}.
The diffusion constants and the critical electric field around which $J$ ceases to be conserved are given in Table~\ref{tab:classical_E_field}.

\begin{table}[]
    \centering
    \caption{The ``diffusion constant'' $\alpha$ and the field strengths around which we expect the electric field to significantly affect the sticking time of the collision complex, $E^\text{traj}_\text{elec}$.
Note that the precise values are subject to ambiguity in the point at which the transition to $J$ non-conservation occurs, which is not a sharp transition, and is also affected by uncertainty in the DIM interaction potentials used. }
    \begin{tabular}{cSS}
        \hline
        \hline
        System&\multicolumn{1}{c}{$\hbar\alpha$}&\multicolumn{1}{c}{$E^\text{traj}_\text{elec}$}\\
        &\multicolumn{1}{c}{($\text{ns}^{-1/2}[\text{V/cm}]^{-1}$)}&\multicolumn{1}{c}{(V/cm)}\\
        \hline
         \multicolumn{1}{c}{RbCs+RbCs}&9.8e-4&0.54\\
         \multicolumn{1}{c}{NaK+NaK}&7.6e-4&7.0\\
         \multicolumn{1}{c}{NaK+K}&4.1e-4&2214\\
         \hline
         \hline
    \end{tabular}

    \label{tab:classical_E_field}
\end{table}

\section{Symmetry breaking in complexes described using random matrix theory}

Next, we consider the role of symmetry breaking in the formalism for describing ultracold collisions proposed by Mayle \textit{et al.}~\cite{mayle:12,mayle:13}.
In this formalism, the long-range physics is described quantum mechanically using multichannel quantum defect theory (MQDT).
The short-range physics, \emph{i.e.}, the chaotic dynamics of the collision complex, is described by random matrix theory (RMT).
In particular, the short-range Hamiltonian $\hat{H}_0$ is drawn from the Gaussian Orthogonal Ensemble (GOE) \cite{mayle:12},
which is fully characterized by the density of states $\rho$.
This gives rise to a short-range reactance matrix
\begin{equation}
    K^\text{SR}=-\frac{1}{2}\mathbf{w}^\dagger(E\mathbf{I}-\mathbf{H_0}-i\Gamma^\mathrm{inel}/2\,\mathbf{I})^{-1}\mathbf{w}, 
    \label{eq:K_matrix}
\end{equation}
where $\mathbf{w}$ is a vector describing the coupling between the resonances and the single open asymptotic channel \cite{mayle:13} and $E$ is the energy.
Since there is only one open scattering channel, $K^\text{SR}$ is a scalar.
The short-range loss rate $\Gamma^\mathrm{inel}$ was introduced by Christianen \textit{et al.}~\cite{christianen:21} to describe losses by processes such as chemical reactions, inelastic scattering, or photoexcitation.
Unless otherwise specified, we will set $\Gamma^\mathrm{inel}=0$.
The vector $\mathbf{w}$ is normally distributed with mean $0$ and standard deviation
\begin{equation}\label{eq:W_matrix}
    \sigma_{\mathbf{w}}=\sqrt{\frac{1}{2\pi\rho}}, 
\end{equation}
determined by the Weisskopf estimate \cite{christianen:21,mitchell:10}.
Neglecting the effects of long-range interactions, that could otherwise be included by the machinery of MQDT, we obtain the $S$-matrix
\begin{equation}
    S^\text{SR}=\frac{1+iK^\text{SR}}{1-iK^\text{SR}}. 
\label{eq:Ssr}
\end{equation}
Short-range loss can be characterized by the parameter
\begin{align}
y = \frac{1-|S^\text{SR}|}{1+|S^\text{SR}|},
\end{align}
which will be nonzero only for $\Gamma^\mathrm{inel}>0$.
From the $S$-matrix we can also calculate the time delay \cite{smith:60}
\begin{align}
    \tau &=-i\hbar S^\mathrm{SR}\frac{dS^{\mathrm{SR}\ \dagger}}{dE} \nonumber \\
         &= \hbar \frac{\mathbf{w}^\dagger \left( E\mathbf{I} -\mathbf{H}_0 \right)^{-2} \mathbf{w} }{1+\left[\frac{1}{2} \mathbf{w}^\dagger \left( E\mathbf{I} -\mathbf{H}_0 \right)^{-1} \mathbf{w} \right]^2},
\end{align}
where we have assumed $\Gamma^\mathrm{inel} = 0$.
The energy derivative is evaluated analytically given the simple form assumed for the short-range reactance matrix.
One of the main results of Mayle \textit{et al.}~\cite{mayle:12,mayle:13} has been that the mean time delay, averaged over an energy window containing many resonances,
is simply the RRKM sticking time, irrespective of long-range effects.

We emphasize that we here describe the effect of a static electric field on the short-range physics only.
This is complimentary to a recent study of the effect of external fields on the long-range physics~\cite{quemener:22}.
The effects on both the long-range and short-range physics can be described in a unified manner~\cite{croft:20,christianen:21} using quantum defect theory.
The description of the short-range Hamiltonian and reactance matrix is unchanged,
but instead of using Eq.~\eqref{eq:Ssr}, the physical $S$-matrix including a quantum mechanical treatment of the long-range physics is obtained as~\cite{christianen:21}
\begin{align}
S^\mathrm{phys} &= \exp(2 i \xi) \frac{1+i C^{-2} [(K^{\mathrm{SR}})^{-1} - \tan\lambda]^{-1}}{1-i C^{-2} [(K^{\mathrm{SR}})^{-1} - \tan\lambda]^{-1}}.
\end{align}
Again, in this work we accounted only for the short-range physics which amounts to using the classical limit of the QDT parameters $\xi=0$, $C^{-2}=1$, and $\tan\lambda=0$~\cite{christianen:21}.

We incorporate the effects of electric fields into this formalism by modifying the Hamiltonian matrix, $\mathbf{H}$, and coupling vector, $\mathbf{w}$, used.
We extend the Hamiltonian by adding blocks to the Hamiltonian for each value of the total angular momentum, $J$, up to some truncation $J_\text{max}$.
The blocks are initially uncoupled, and each block is drawn from GOE with a density of states $\rho_J = (2J+1)\rho_0$.
These blocks are then coupled by blocks $\mathbf{C}_{J,J+1}$, which describe the coupling between the different blocks $\mathbf{H}_J$ due to the electric field.
In the field-free eigenbasis, the matrix elements in these blocks are normally distributed random variables with zero mean.
Their variance determined by the Feingold-Peres formula, as described in Sec.~\ref{sec:feingold_peres} below.
The total short-range Hamiltonian is
\begin{equation}
    \mathbf{H}=\begin{pmatrix}
        \mathbf{H}_0&\mathbf{C}_{0,1}&0&\dots&0\\
        \mathbf{C}_{0,1}^\dagger &\mathbf{H}_1&\ddots&&\vdots\\
        0&\ddots&\ddots&&0\\
        \vdots&&&\mathbf{H}_{J_\text{max}-1}&\mathbf{C}_{J_\text{max}-1,J_\text{max}}\\
        0&\dots&0&\mathbf{C}_{J_\text{max}-1,J_\text{max}}^\dagger &\mathbf{H}_{J_\text{max}}\\
    \end{pmatrix}. 
\label{eq:Hsr}
\end{equation}
This block structure results from the dipole selection rule that states with $J$ and $J'$ are coupled only if $J'=J\pm 1$, for $M_J=M_J'=0$.
Dipole transition moments furthermore exist only between opposite parity states.
Since we start in $J=0$ with even parity, it is understood that the parity of the coupled blocks is given by $(-1)^J$.
We assume the long range only couples to the $J=0$ states, so the coupling vector, $\mathbf{w}$ from Eq.~\eqref{eq:K_matrix}, is $\mathbf{w}=(\mathbf{w}_0^T,0,\ldots,0)^T$, where $\mathbf{w}_0$ is the coupling vector of the system with Hamiltonian $\mathbf{H}_0$ describing states with $J=0$, see Eq.~\eqref{eq:W_matrix}. 

\subsection{Calculating the transition dipole moment} \label{sec:feingold_peres}

To compute the magnitude of the coupling $\mathbf{C}_{J,J+1}$ between $J$ and $J+1$ total angular momentum blocks,
we need to calculate the variance of the off-diagonal elements of the dipole operator in the energy eigenbasis.
On average, a squared dipole matrix element is given by
\begin{align}
|\langle\alpha,J|\hat{\mathbf{d}}_x|\beta,J+1\rangle|^2=\frac{S_{\mathbf{d}}(E_\alpha,\omega)}{ \hbar\rho_0\ 4} \left[\frac{1}{2J+1} + \frac{1}{2J+3}\right],
\label{eq:var_autocorr}
\end{align}
which is essentially the Feingold-Peres formula \cite{feingold:86,leitner:94,prosen:94,wilkinson:87,hortikar:98,zare:88,gordon:68b},
but with additional numerical factors that account for the fact that non-zero dipole matrix elements are located in the blocks that connect $J$ with $J-1$ and $J$ with $J+1$ eigenstates. 
The quantity 
$S_\mathbf{d}(E_\alpha,\omega)$ is the Fourier transform of the dipole autocorrelation function
\begin{equation}
    S_{\mathbf{d}}(E_\alpha,\omega)=\frac{1}{2\pi}\int_{-\infty}^{\infty}dt e^{i\omega t}\langle\mathbf{d}(\mathbf{q}_t)\cdot\mathbf{d}(\mathbf{q}_0)\rangle_{\alpha}/3. 
    \label{eq:fourier_autocorrelation}
\end{equation}
Here $\mathbf{d}(\mathbf{q})$ is the dipole moment at coordinates $\mathbf{q}$, $\mathbf{q}_t$ indicates the coordinates at time $t$ of the classical trajectory starting at $\mathbf{q}_0$, and $\langle\cdot\rangle_\alpha$ denotes a phase-space average at energy $E_\alpha$, which is equivalent to a micro-canonical ensemble average.
Hence, the phase-space average is evaluated by averaging over many trajectories.
The ensemble average is also equivalent to a time average, which is used to improve statistics by averaging over the full simulations.
We use the static limit of the autocorrelation function, $S_\mathbf{d}(E_0,\omega)\approx S_\mathbf{d}(E_0,0)$,
which is justified as the typical transition frequency between chaotic eigenstates $\omega = \mathcal{O}(1/\hbar\rho)$ is small compared to the typical frequency of the autocorrelation function, $\omega_\mathrm{ACF} = \mathcal{O}(1/\tau_\mathrm{ACF})$,
or equivalently, that the timescale of the autocorrelation function, $\tau_\mathrm{ACF}$, is short compared to the sticking time $\tau_\mathrm{stick} = 2\pi\hbar\rho$.
We furthermore assume that the dipole autocorrelation function is independent of $J$, which seems reasonable for low $J$ values.

We can now determine the variance of the Stark coupling matrix elements in the $\mathbf{C}_{J,J+1}$ blocks 
\begin{equation}
    \text{Var}[(\mathbf{C}_{J,J+1})]=\frac{E_\text{elec}^2 S_{\mathbf{d}}(E_\alpha,\omega)}{ \hbar\rho_0\ 4} \left[\frac{1}{2J+1} + \frac{1}{2J+3}\right],
\end{equation}
where $E_\text{elec}$ is the electric field.
Following Leitner \textit{et al.}~\cite{leitner:94}, we define a dimensionless parameter
\begin{align}
\Omega = (\rho_0 + \rho_1 ) \sqrt{\text{Var}[(\mathbf{C}_{J,J+1})},
\end{align}
which is the root mean square coupling between $J=0$ and $J=1$ states in units of their mean level spacing, $1/(\rho_0+\rho_1)$.
This is essentially the square root of the parameter proposed by Leitner \textit{et al.}~\cite{leitner:94}, for a different Hamiltonian with a different block structure.
Taking the square root has the advantage that our parameter $\Omega$ is interpreted as a coupling strength, and hence scales with the electric field $E_\text{elec}$.
Angular momentum is conserved for $\Omega\ll 1$, and angular momentum is completely scrambled for $\Omega \gg 1$.
Next, we investigate the finite $\Omega$ for which the transition between these limits occurs.

We ran 1000 (or 600 for NaK+K) trajectories for 10 ns and recorded the electric dipole moment (every 4 time steps for the diatom-diatom systems; every 160 steps for NaK+K) in zero field.
We then calculated the dipole autocorrelation functions, shown in Fig.~2 of the main text.
Before calculating the autocorrelation function of a trajectory we subtracted the square of the average value of the dipole moment during that trajectory.
This ensured that the autocorrelation approaches zero at late times.
The static limits of the dipole autocorrelation functions, $S_\mathbf{d}(E_0,0)$, are given in Table \ref{tab:S_P}.
This table also includes a critical electric field
\begin{align}
E_\text{elec}^\mathrm{RMT} = \frac{1}{2\sqrt{S_\mathbf{d}(E_0,0) \rho_0}}
\end{align}
which is the electric field strength at which $\Omega = 1$.
We note that the critical electric fields estimated in this way are roughly in agreement with the electric fields obtained by extrapolating angular momentum in trajectory simulations, $\mathrm{E}_\mathrm{elec}^\mathrm{traj}$ in Table~\ref{tab:classical_E_field}.

\begin{table}
    \centering
    \caption{Static limit of the dipole autocorrelation function, $S_\mathbf{d}(E_0,0)$, and the critical electric field $E_\text{elec}^\text{RMT}$ at which $\Omega=1$, the point around which the transition to $J$ non-conservation occurs.
Note that the precise values are subject to ambiguity in the position of the transition, which is not sharp, and also affected by uncertainty in the DIM interaction potentials used. }
    \begin{tabular}{lSSSS}
    \hline
    \hline
    System&\multicolumn{1}{c}{$S_\mathbf{d}(E_0,0)$}&\multicolumn{1}{c}{$E_\text{elec}^\text{RMT}$}&\\
    &\multicolumn{1}{c}{($e^2a_0^2\hbar E_h^{-1}$)}&\multicolumn{1}{c}{(V/cm)}\\
    \hline
    \multicolumn{1}{c}{RbCs+RbCs}&39264&3.7\\
    \multicolumn{1}{c}{NaK+NaK}&23860&48\\
    \multicolumn{1}{c}{NaK+K}&7435&14763\\
    \hline
    \hline
    \end{tabular}

    \label{tab:S_P}
\end{table}

\subsection{Transition between angular momentum conservation and non-conservation in RMT}

The presence of an electric field causes the total angular momentum of the collision complex to be no longer conserved.
We investigate this transition for finite $\Omega$ by observing the effect of symmetry breaking on three different properties;
the energy level statistics, the distribution of time delays, and the short-range loss parameter.

\subsubsection{Level statistics}

\begin{figure}
\centering
\includegraphics[width=0.475\textwidth]{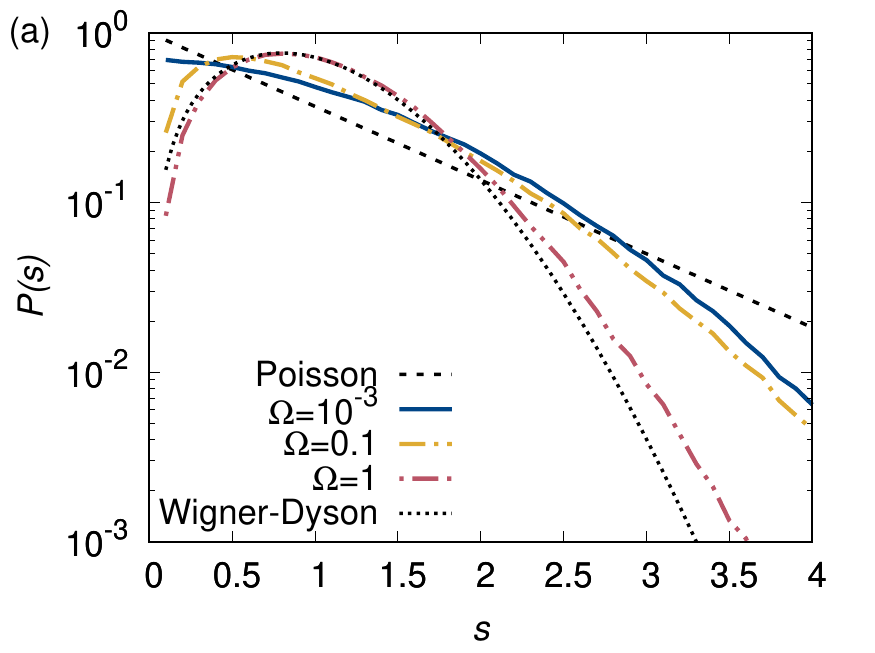}
\includegraphics[width=0.475\textwidth]{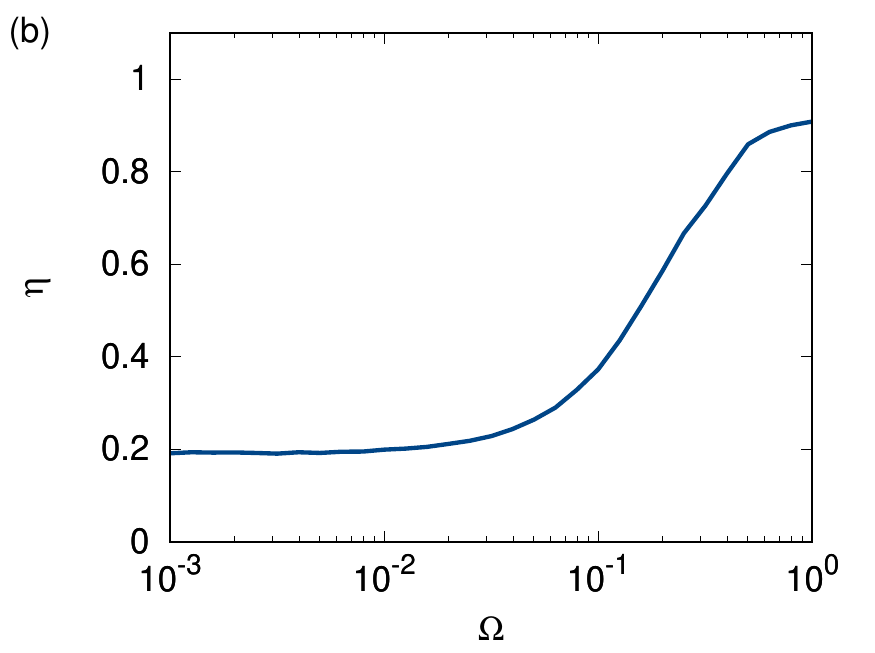}
\caption{
Panel {\bf (a)} shows the probability distribution of nearest-neighbor energy spacings in units of the mean energy spacing, $s$, for $J_\mathrm{max} = 3$ and various $\Omega$.
For low $\Omega$, there is energy-level-repulsion between eigenstates of the same $J$ only, leading to a distribution close to Poissonian.
For high $\Omega$, the total system becomes ergodic as $J$ is no longer conserved, and the distribution of nearest-neighbor energy spacings approaches the Wigner-Dyson distribution.
Panel {\bf (b)} shows results of fitting the distribution of nearest-neighbor energy spacings with a Brody distribution, which interpolates between Poisson at $\eta=0$ and Wigner-Dyson at $\eta=1$,
illustrating $J$ ceases to be conserved around $\Omega=1$.
}
\label{fig:levelstatistics}
\end{figure}

We first analyze the symmetry-breaking transition by analyzing the energy level statistics of $\mathbf{H}$.
To this end, we diagonalize $\mathbf{H}$ and keep the center 1/16th of the spectrum, in order to avoid artifacts due to the edges of the spectrum.
We then determine the spacings between subsequent eigenvalues in units of the mean spacing,
\begin{align}
s_i &= (E_{i+1} - E_i) \rho_\mathrm{tot}, \nonumber \\
\rho_\mathrm{tot} &= \sum_{J=0}^{J_\mathrm{max}} \rho_J = \rho_0 \sum_{J=0}^{J_\mathrm{max}} (2J+1) = (1+J_\mathrm{max})^2 \rho_0.
\end{align}
The distribution of energy spacings is shown in Fig.~\ref{fig:levelstatistics}.
For low $\Omega$, there is energy-level-repulsion between eigenstates of the same $J$ only, leading to a distribution close to Poissonian.
For high $\Omega$, the total system becomes ergodic as $J$ is no longer conserved, and the distribution of nearest-neighbor energy spacings approaches the Wigner-Dyson distribution.
In order to quantitatively follow this symmetry-breaking transition we fit the distribution of energy levels with the Brody distribution~\cite{brody:73}
\begin{align}
P_\eta (s) &= c_\eta (1+\eta) s^\eta \exp\left(-c_\eta s^{\eta+1}\right), \nonumber \\
c_\eta &= \Gamma\left[\frac{\eta+2}{\eta+1}\right]^{\eta+1},
\end{align}
which interpolates between the Poisson distribution at $\eta=0$ and the Wigner-Dyson distribution at $\eta=1$.
Figure~\ref{fig:levelstatistics} shows this transition occurs roughly at $\Omega=1$.

\subsubsection{Time delay distribution}

Since the level statistics may not be directly observable in experiments,
we consider additional measures of the transition to $J$ non-conservation.
In particular, we next consider the time delay in $s$-wave collisions.
Mayle \emph{et al.}~\cite{mayle:13} have shown that the time delay, when averaged over many resonances,
is sensitive only to the density of resonances and not their widths.
This means that any finite coupling between different $J$ states will immediately increase an energy-averaged time delay,
provided the energy resolution is sufficient to sample the very narrow resonances caused by minute couplings between $J$ states.
Therefore, we instead consider the \emph{distribution} of time delays, rather than its energy average.
We probe this distribution by repeatedly computing the time delay for different random realizations of the statistical short-range Hamiltonian.

\begin{figure}
\centering
\includegraphics[width=0.5\textwidth]{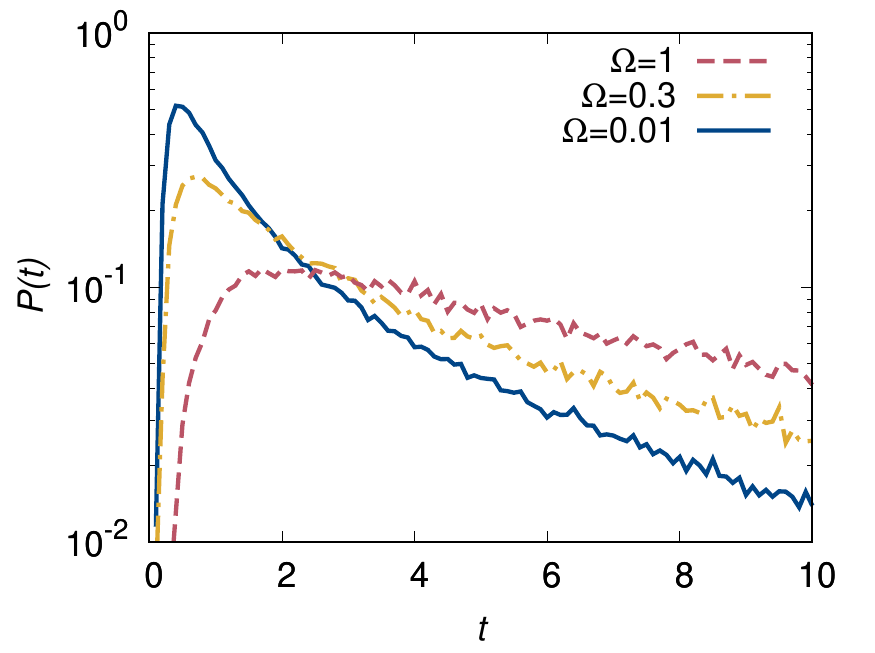}
\caption{
Distribution of time delays for $J_\mathrm{max} = 2$ and various $\Omega$.
For increasing $\Omega$, the distribution stretches to longer time delays as $J$ is no longer conserved and the effective density of states increases.
}
\label{fig:timedelayLambda}
\end{figure}

\begin{figure}
\centering
\includegraphics[width=0.475\textwidth]{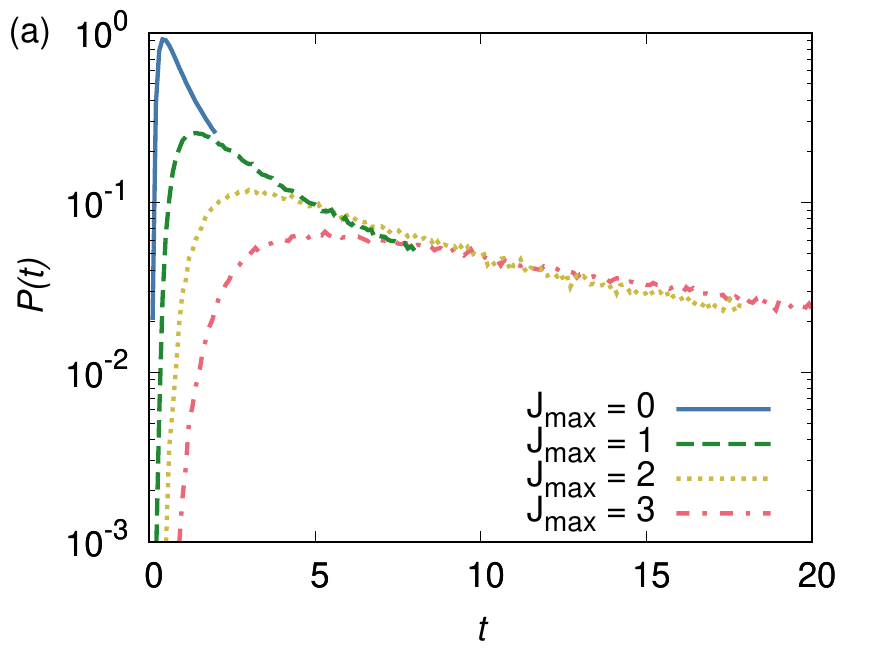}
\includegraphics[width=0.475\textwidth]{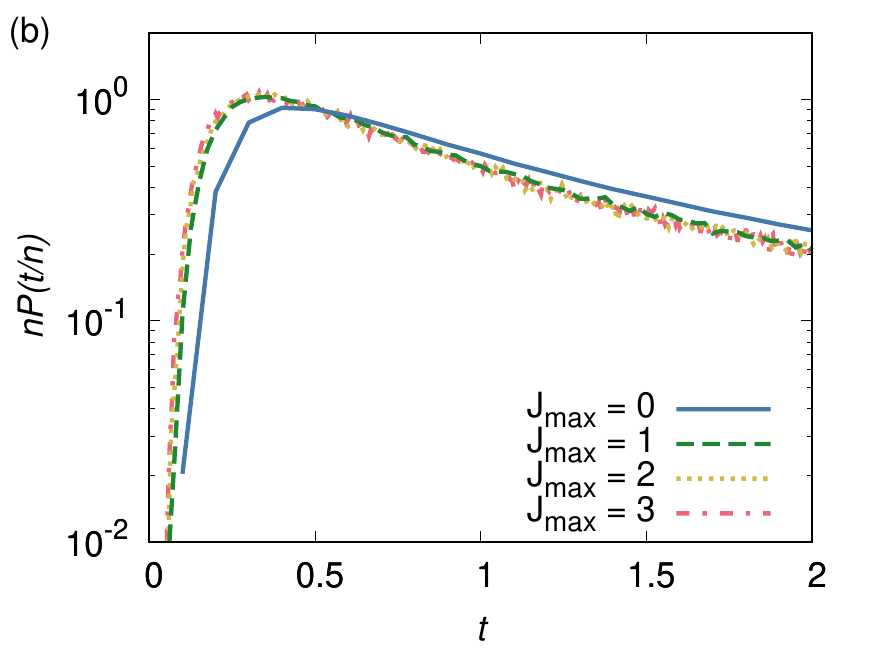}
\caption{
Distribution of time delays for fixed $\Omega=4$, so that $J$ is no longer conserved, for various $J_\mathrm{max}$.
Panel {\bf (a)} shows the distribution stretches to longer times for increasing $J_\mathrm{max}$.
This occurs because the \emph{total} density of states increases with $n=(1+J_\mathrm{max})^2$.
Panel {\bf (b)} shows that the distributions collapse onto one another by rescaling with the total density of states.
This is the expected behavior for chaotic dynamics, as the statistical properties of the Hamiltonian matrix are determined by the density of states alone.
}
\label{fig:timedelayJ}
\end{figure}

\begin{figure}
\centering
\includegraphics[width=0.5\textwidth]{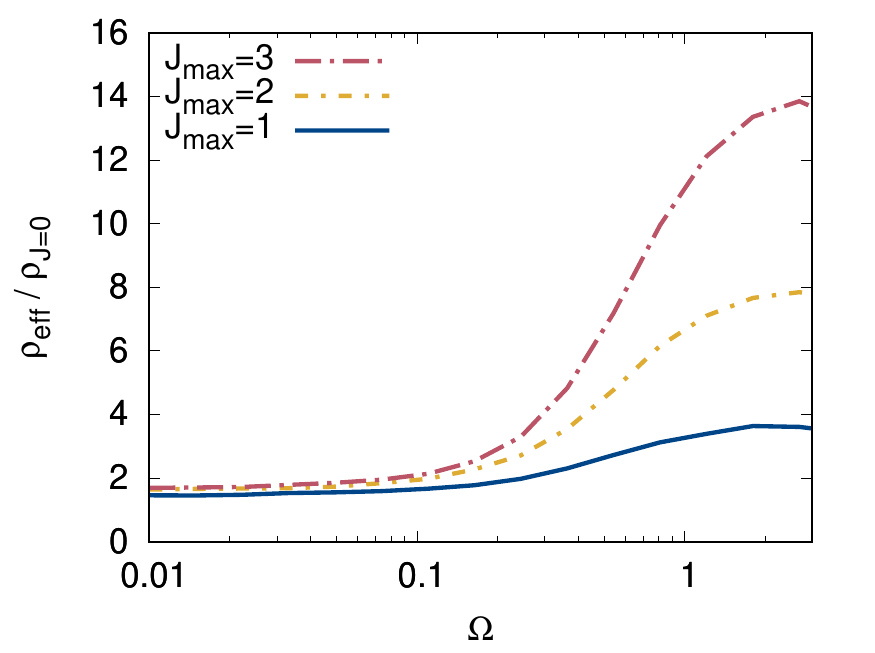}
\caption{
Effective density of states $\rho_\mathrm{eff} = n \rho_{J=0}$ determined by fitting the distribution of time delays, $P(t)$, with a stretched $J=0$ distribution, $P(nt)/n$, where $n$ is the fitting parameter.
The effective density of states approaches $\rho_{J=0}$ at low $\Omega$, and increases approximately $(1+J_\mathrm{max})^2$-fold for large $\Omega$, approaching the total density of states.
Although the total density of states depends on $J_\mathrm{max}$, the transition universally occurs around $\Omega=1$.
}
\label{fig:timedelayRho}
\end{figure}

Figure~\ref{fig:timedelayLambda} shows the distribution of time delays for $J_\mathrm{max}=2$ for several values of $\Omega$,
showing that for increasing $\Omega$, the distribution stretches to longer time delays as $J$ is no longer conserved and the effective density of states increases.
The idea is that when $\Omega=0$, the distribution is determined only by the chaotic $J=0$ states,
whereas at high $\Omega$ the dynamics is thought to become chaotic even in $J$.
This means that at both extremes, the distribution of time delays is determined by coupling to a set of chaotic states described by a Gaussian Orthogonal Ensemble.
Hence, these sets of states are statistically equivalent, and all that changes is the density of states that characterizes the set of chaotic states.
That is, we expect the distribution of time delays is identical except that the time scale $\tau \propto \hbar \rho$ has increased.
Figure~\ref{fig:timedelayJ} confirms this picture,
showing that at $\Omega=4$ the distribution of time delays stretches by a factor $(1+J_\mathrm{max})^2$.
At finite $\Omega$, it is not entirely clear that the distribution of time delays will have the same form, but for simplicity we make this assumption.
For each $\Omega$ we perform a nonlinear least squares fit of the distribution obtained numerically to the function $P_0(nt)/n$,
where $P_0(t)$ is the distribution for $J=0$.
The fitting parameter $n$ defines the effective density of states as
\begin{align}
\rho_\mathrm{eff} = n \rho_{0}.
\end{align}
Figure~\ref{fig:timedelayRho} shows that the effective density of states smoothly transitions around $\Omega=1$ from the $J=0$ density of states, $\rho_0$, to the total density of states, $(1+J_\mathrm{max})^2\rho_0$.

\subsubsection{Short-range loss}

\begin{figure}
\centering
\includegraphics[width=0.5\textwidth]{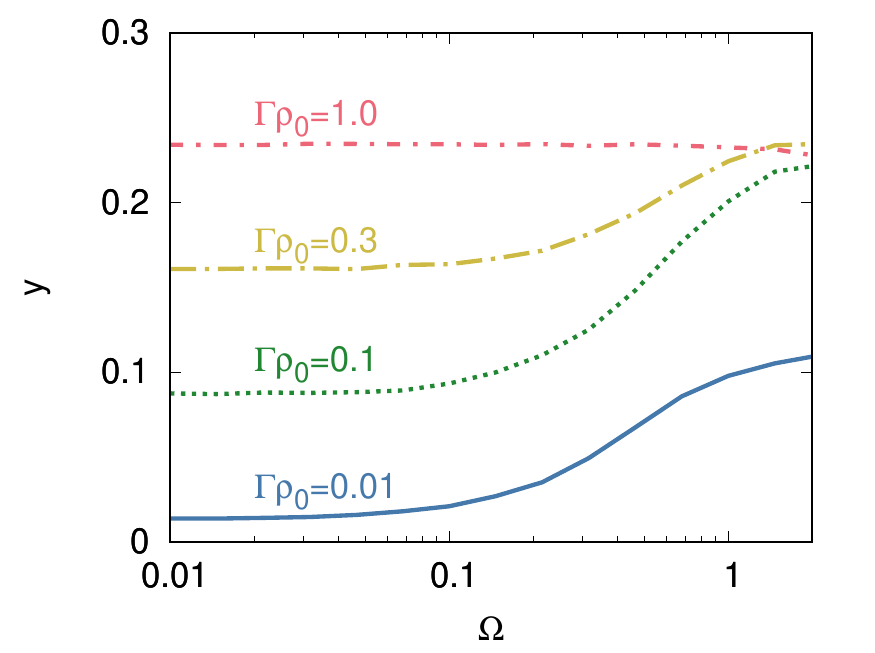}
\caption{
Short-range loss parameter $y$ as a function of $\Omega$ for various short-range loss rates $\Gamma^\mathrm{inel}$ and fixed $J_\mathrm{max}=3$.
For overlapping resonances, $\Gamma^\mathrm{inel} \rho_\mathrm{eff} \gg 1$, the loss parameter approaches $y=1/4$.
For $\Gamma^\mathrm{inel} \rho_0 = 10$, this is the case for all $\Omega$.
For lower $\Gamma^\mathrm{inel} \rho_0$, however, at low $\Omega$ the effective density of states is just the $J=0$ density of states, hence $\Gamma^\mathrm{inel} \rho_\mathrm{eff} < 1$ such that the mean loss parameter is smaller.
As $\Omega$ is increased to unity, the effective density of states increases by a factor $(1+J_\mathrm{max})^2=16$, such that $y$ approaches $1/4$.
This occurs universally around $\Omega=1$, regardless the precise value of $\Gamma^\mathrm{inel}$ or $\rho_\mathrm{total}$, as long as $\Gamma^\mathrm{inel} \rho_\mathrm{total} \gg 1$.
For the numerical results shown, this occurs except for the case $\Gamma^\mathrm{inel} \rho_0 = 0.01$.
Here, the effective density of states does grow 16-fold to the \emph{total} density of states around $\Omega = 1$,
but this is insufficient to reach the limit of overlapping resonances for $\Gamma^\mathrm{inel} \rho_0 = 0.01$ and $J_\mathrm{max}=3$.
}
\label{fig:loss}
\end{figure}

Finally, we consider observing the same transition to $J$ non-conservation in collisional loss.
Following Ref.~\cite{christianen:21}, short-range loss is characterized by a loss parameter $y$,
which can be determined from $y = (1-|S^\mathrm{SR}|)/(1+|S^\mathrm{SR}|)$.
We determine the short-range $S$-matrix as above,
but add an imaginary component $-i\Gamma^\mathrm{inel}/2$ to the energy of the resonance states.
The parameter $\Gamma^\mathrm{inel}$ determines the rate of loss of the collision complex,
and in the limit of fast loss $\Gamma^\mathrm{inel} \rho \gg 1$ one obtains $y=1/4$~\cite{christianen:21}.
Figure~\ref{fig:loss} shows the resulting loss parameter as a function of $\Omega$ for several $\Gamma^\mathrm{inel}$ obtained numerically for $J_\mathrm{max}=3$.
Where $\Gamma^\mathrm{inel} \rho_0 \gg 1$, we obtain $y=1/4$ regardless the electric field $\Omega$.
For lower $\Gamma^\mathrm{inel} \rho_0$ the loss parameter at $\Omega=0$ is reduced.
However, as we increase $\Omega$ the effective density of states increases as $J$ is no longer conserved.
The total density of states for $J_\mathrm{max}=3$ is $16\rho_0$,
which puts us in the regime of overlapping resonances and results in $y=1/4$, for all loss parameters shown except $\Gamma^\mathrm{inel} \rho_0 = 0.01$.
Again, we observe that the transition to $J$ non-conservation occurs around $\Omega =1$.

\section{Black-body radiation, spontaneous emission, and monochromatic radiation}

We also studied the loss rate of collision complexes due to the absorption of black-body radiation, spontaneous emission, and absorption of electromagnetic radiation.
The total transition rate from energy level $\alpha$ due to absorption of black body radiation \cite{buhmann:08} is given by
\begin{equation}
    \Gamma^{(T)}_\alpha=\sum_\beta \frac{|\omega_{\alpha,\beta}|^3|\langle\alpha|\hat{\mathbf{d}}|\beta\rangle|^2}{3\pi\hbar\epsilon_0c^3}f_\text{Boltz,T}(|\omega_{\alpha,\beta}|), 
\label{eq:bb}
\end{equation}
where we sum over all other energy levels, $\beta$, we have $\hbar\omega_{\alpha,\beta} = E_\beta-E_\alpha$, $\epsilon_0$ is the vacuum permittivity, $c$ is the speed of light, $f_\text{Boltz,T}(\omega)=1/\{\exp[\hbar\omega/(k_BT)]-1\}$ is the Boltzmann distribution, $k_B$ is the Boltzmann constant, and $T$ is the temperature of the environment, not the molecules. The rate of spontaneous emission is
\begin{equation}
    \Gamma^{(0)}_\alpha=\sum_{E_\beta<E_\alpha} \frac{|\omega_{\alpha,\beta}|^3|\langle\alpha|\hat{\mathbf{d}}|\beta\rangle|^2}{3\pi\hbar\epsilon_0c^3}. 
\label{eq:se}
\end{equation}
The probability of transitioning out of the initial state after time $t$ due to absorption of monochromatic radiation is
\begin{equation}
    \Gamma^{(\omega)}_\alpha=\sum_\beta\frac{\pi u(\omega_{\alpha,\beta}) |\langle\alpha|\hat{\mathbf{d}}|\beta\rangle|^2} {3\hbar^2\epsilon_0}, 
\end{equation}
with $u(\omega)$ the energy density per unit frequency at frequency $\omega$ \cite{bransden:03}.

We replaced the sums over states above by integrals over the energy and used the Feingold-Peres formula Eq.~\eqref{eq:var_autocorr} for the transition dipole moment.
The resulting expressions are
\begin{align}
    \Gamma^{(T)} &= \hbar \int_{-\infty}^\infty \frac{|\omega|^3|S_\mathbf{d}(0,\omega)}{\pi\hbar\epsilon_0c^3}f_\text{Boltz,T}(|\omega|)d\omega ,\label{eq:integrand1} \\
    \Gamma^{(0)} &= \hbar \int_{-\infty}^{0} \frac{|\omega|^3S_\mathbf{d}(0,\omega)}{\pi\hbar\epsilon_0c^3}d\omega, \label{eq:integrand2}\\
    \Gamma^{(\omega)} &= \frac{\pi S_\mathbf{d}(0,\omega)}{\hbar^2\epsilon_0}u(\omega). 
\end{align}
The integrands in Eqs.~\eqref{eq:integrand1} and \eqref{eq:integrand2} are denoted $\tilde{(\Gamma)}^T$, $\tilde{(\Gamma)}^0$, respectively,
and these are shown in Fig.~4 of the main text.
The Fourier transform of the dipole autocorrelation function is appreciable only for small values of $\omega$,
while the factor $\omega^3$ in Eqs.~\eqref{eq:bb} and \eqref{eq:se} emphasizes the contribution of higher transition frequencies for black-body heating and spontaneous emission.
Hence, the behavior in the high-frequency wing of $S(E_0, \omega)$ is important.
We cut off the autocorrelation function after $0.32$~ns and multiplied the result by the Hann window function \cite{press:07} before applying the Fourier transform,
in an effort to reduce noise in the high-frequency tail of $S(E_0,\omega)$ \cite{harris:78}. 

Numerical results for the loss rate due to spontaneous emission and the absorption of black-body radiation are shown in Fig.~\ref{fig:loss_rates_T}.
The loss rate due to both loss processes is slower than 1~s$^{-1}$, and plays no role during a typical sticky collision.
We also plotted the loss rate due to a monochromatic radiation field with frequency $\omega$ ($\Gamma^\omega$) divided by the energy density, $u(\omega)$, of the field, see Fig.~\ref{fig:loss_due_to_field}.

\begin{figure}
    \centering
    \includegraphics[width=0.475\textwidth]{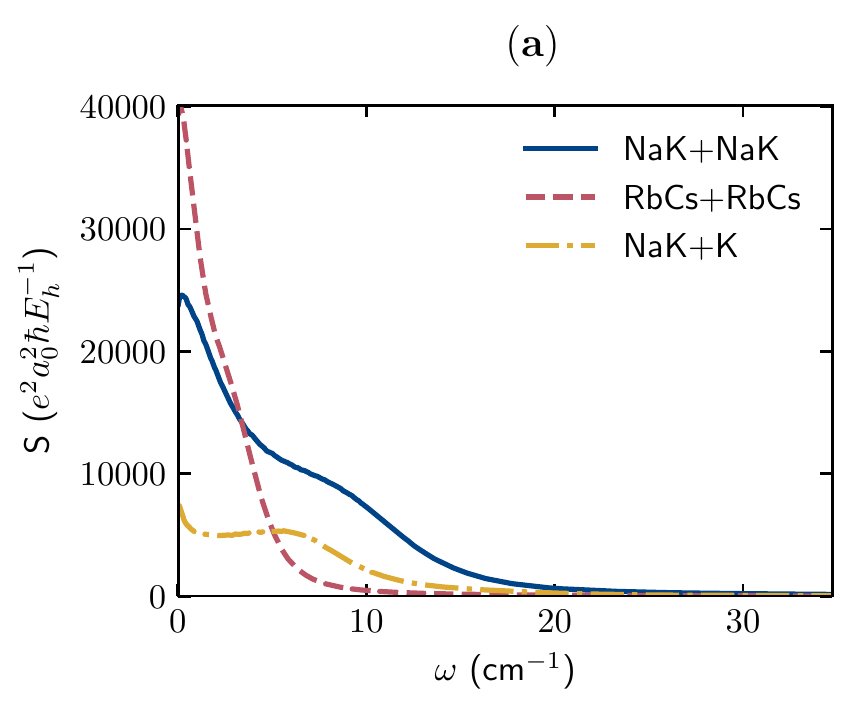}
    \includegraphics[width=0.475\textwidth]{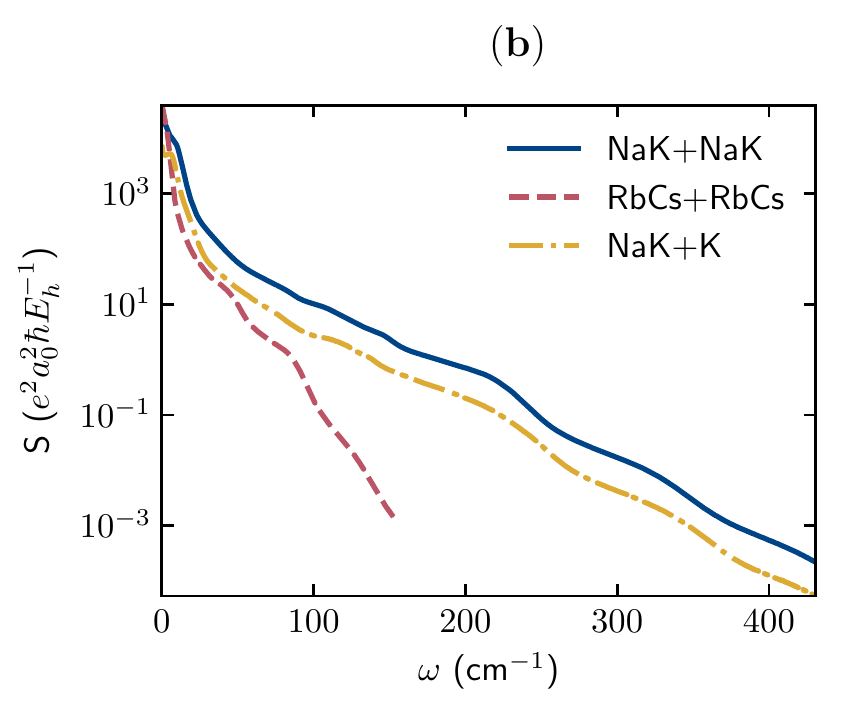}
    \caption{Fourier transform of the dipole autocorrelation function, $S_\mathbf{d}(E_0, \omega)$, on \textbf{(a)} linear and \textbf{(b)} logarithmic scale.
    \label{fig:Fourier_autocorrelation_dipole}}
\end{figure}

\begin{figure}
    \centering
    \includegraphics{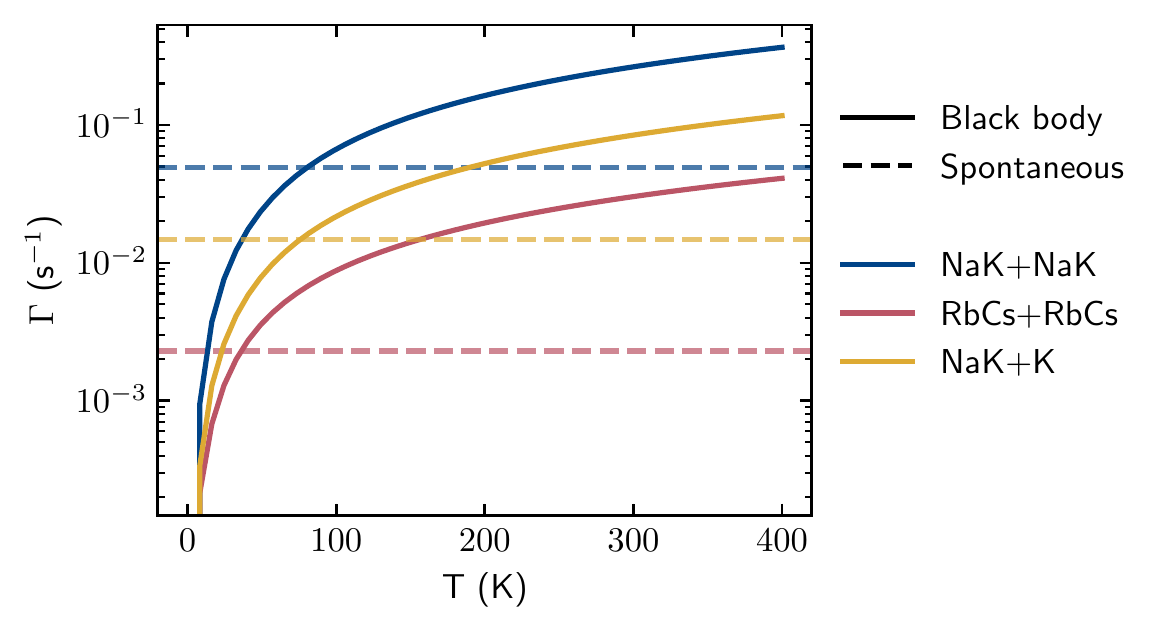}
    \caption{Loss rates due to spontaneous emission, $\Gamma^{(0)}$ (dashed horizontal lines), and transitions induced by black-body radiation at temperature $T$, $\Gamma^{(T)}$ (solid curved lines). }
    \label{fig:loss_rates_T}
\end{figure}

\begin{figure}
    \centering
    \includegraphics[width=0.7\textwidth]{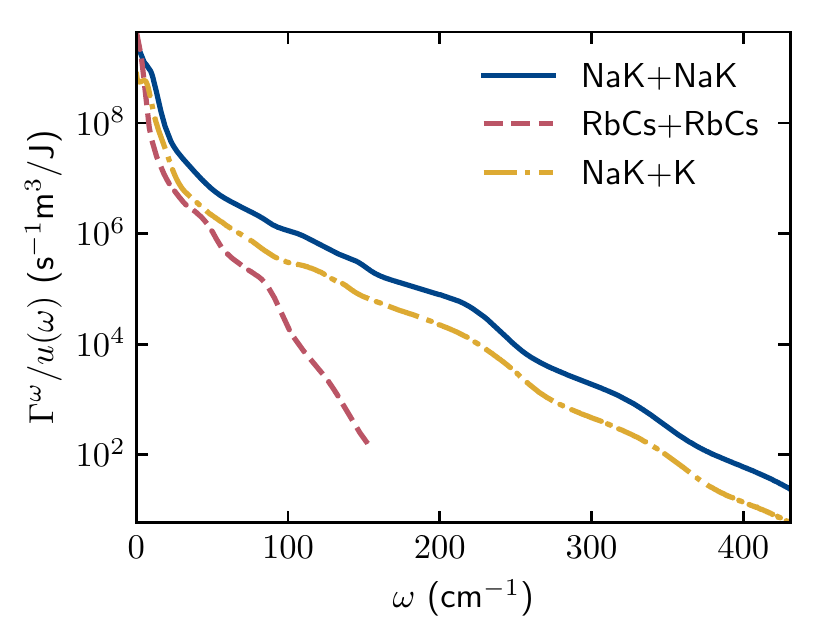}
    \caption{Loss rate, $\Gamma^{(\omega)}$, due to absorption of monochromatic radiation at frequency $\omega$, normalized to the energy density of the radiation field.
    \label{fig:loss_due_to_field}}
\end{figure}

\section{Hyperfine coupling}

In this section we consider the non-conservation of the nuclear spin state during a sticky collision.
To simplify the discussion, we consider only the strongest hyperfine coupling,
which is the interaction between the nuclear quadrupole moment and the electric field gradient.
This interaction takes the form~\cite{aldegunde:17}
\begin{align}
\hat{H}_\mathrm{eQq} = (eQq) \frac{\sqrt{6\pi}}{2i(2i-1)} \left[ \left[ \hat{i} \otimes \hat{i} \right]^{(2)} \otimes Y^{(2)}(\hat{r}) \right]^{(0)}_0.
\label{eq:hf}
\end{align}
Here, $(eQq)$ is the coupling constant that we take from the literature as discussed below,
$\hat{i}$ is the nuclear spin angular momentum operator, a rank-1 tensor with spherical components $\hat{i}_0 = \hat{i}_z$ and $\hat{i}_{\pm 1} = \mp ( \hat{i}_x \pm \hat{i}_y)/\sqrt{2}$,
and $Y^{(2)}(\hat{r})$ is a rank-2 tensor with spherical components that are the density-normalized spherical harmonics, $Y_{2,m}(\hat{r})$, depending on the polar angles of the molecular axis $\hat{r}$.
The symbol
\begin{align}
\left[ \hat{A}^{(k_1)} \otimes \hat{B}^{(k_2)} \right]^{(k)}_q = \sum_{q_1,q_2} \hat{A}_{k_1,q_1} \hat{B}_{k_2,q_2} \langle k_1 q_1 k_2 q_2 | k q \rangle
\end{align}
is the spherical component $q$ of the irreducible rank-$k$ tensor product of two tensors, $\hat{A}$ and $\hat{B}$ of rank $k_1$ and $k_2$ respectively,
and $\langle k_1 q_1 k_2 q_2 | k q \rangle$ is a Clebsch-Gordan coefficient.
This is the form of the coupling for a free molecule, and for simplicity we assume this also holds in the collision complex.
This interaction couples the nuclear spin to the molecular axis.
The physical picture is that the internal dynamics of the collision complex reorients the molecular axes,
resulting in a fluctuating Hamiltonian for the nuclear spin degrees of freedom, which can result in transitions between different spin states.

To simplify the discussion further, we consider only a single nuclear spin per molecule.
For $^{87}$Rb$^{133}$Cs+$^{87}$Rb$^{133}$Cs collision complexes we consider the Rb $i=3/2$ nuclear spin, for which the coupling constant $(eQq) \approx -779$~kHz is larger than the 45~kHz for Cs.
For $^{23}$Na$^{40}$K+$^{23}$Na$^{40}$K we consider the K $i=4$ nuclear spin as this has the larger coupling constant, $(eQq) \approx 899$~kHz versus $-187$~kHz for Na~\cite{aldegunde:17}.

Next, we assume we start in the state $m_i=i$.
The hyperfine coupling considered here will couple this state to states with $m_i=i-1$ and $i-2$.
This selection rule results from the second-rank coupling.
The relevant non-zero matrix elements of Eq.~\eqref{eq:hf} in the basis $|i,m_i\rangle$ are
\begin{align}
\langle ^3/_2, ^3/_2 | \hat{H}_\mathrm{eQq} | ^3/_2,  ^1/_2 \rangle &= -(eQq) \sqrt{\frac{\pi}{10}} Y_{2,-1}(\hat{r}), \nonumber \\
\langle ^3/_2, ^3/_2 | \hat{H}_\mathrm{eQq} | ^3/_2, -^1/_2 \rangle &= (eQq) \sqrt{\frac{\pi}{10}} Y_{2,-2}(\hat{r}), \nonumber \\
\langle 4, 4 | \hat{H}_\mathrm{eQq} | 4,3 \rangle &= -(eQq)  \sqrt{\frac{3\pi}{80}} Y_{2,-1}(\hat{r}), \nonumber \\
\langle 4, 4 | \hat{H}_\mathrm{eQq} | 4,2 \rangle &=  (eQq)  \sqrt{\frac{3\pi}{280}} Y_{2,-2}(\hat{r}).
\label{eq:hf_matel}
\end{align}
For the ``mechanical'' angular momentum, there similarly exists a selection rule $|\Delta J| \le 2$.
This follows from the rank of the spherical harmonic.
As a result, the $J=0$ initial state is coupled to final $J=2$ states only.

\begin{figure}
    \centering
    \includegraphics[width=0.475\textwidth]{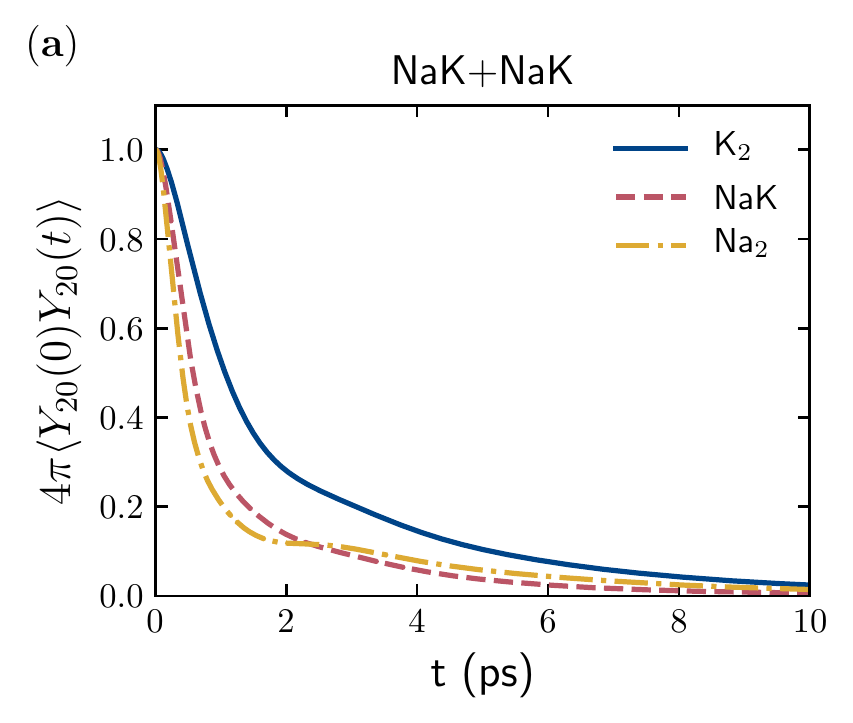}
    \includegraphics[width=0.475\textwidth]{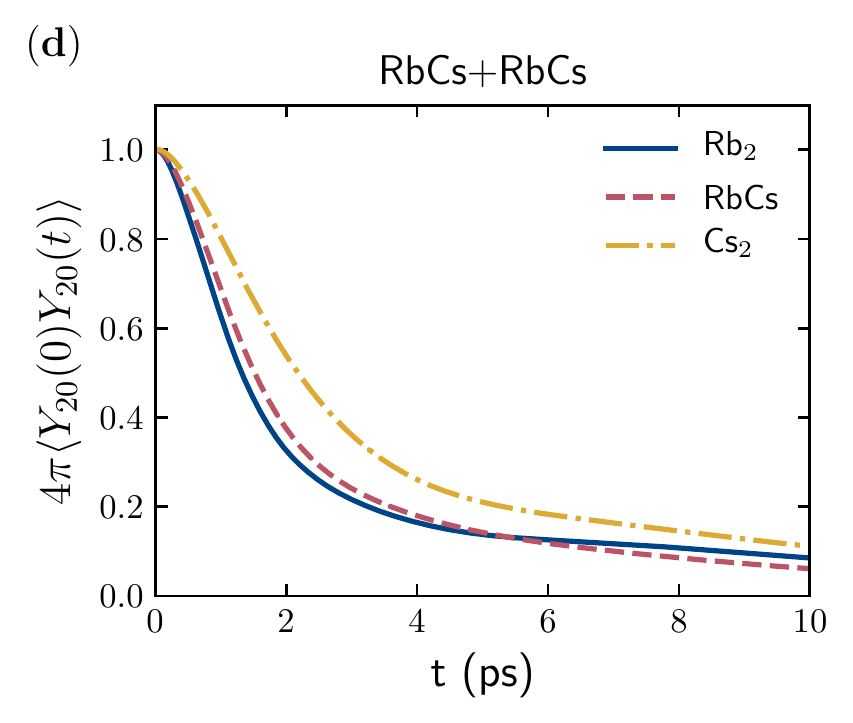}
    \includegraphics[width=0.475\textwidth]{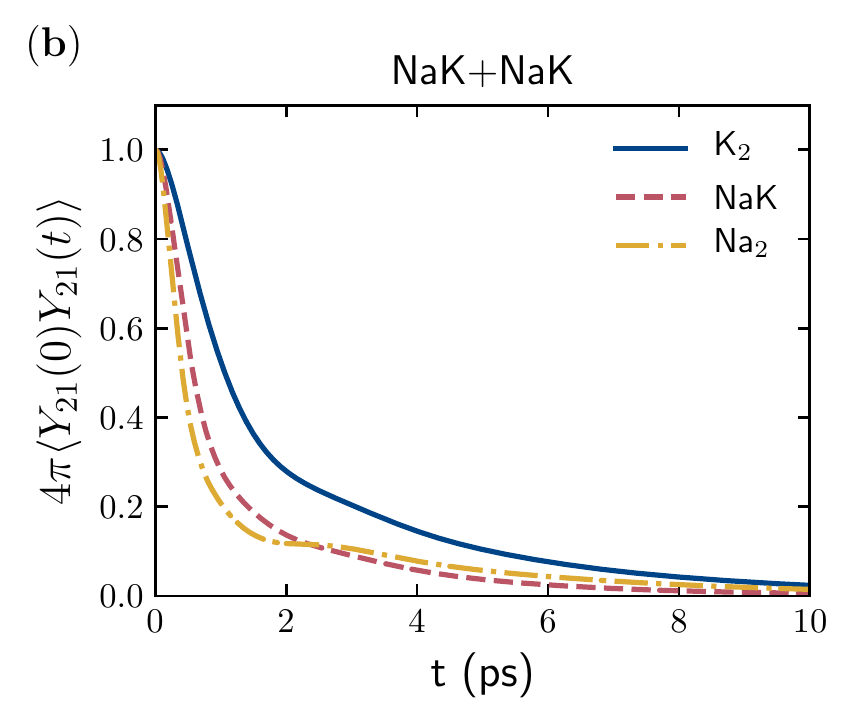}
    \includegraphics[width=0.475\textwidth]{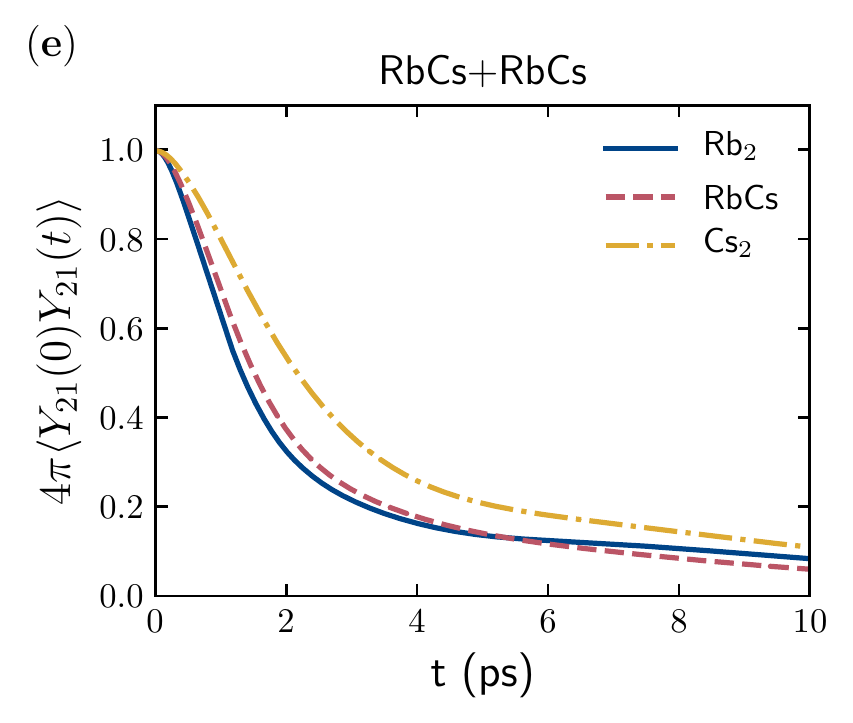}
    \includegraphics[width=0.475\textwidth]{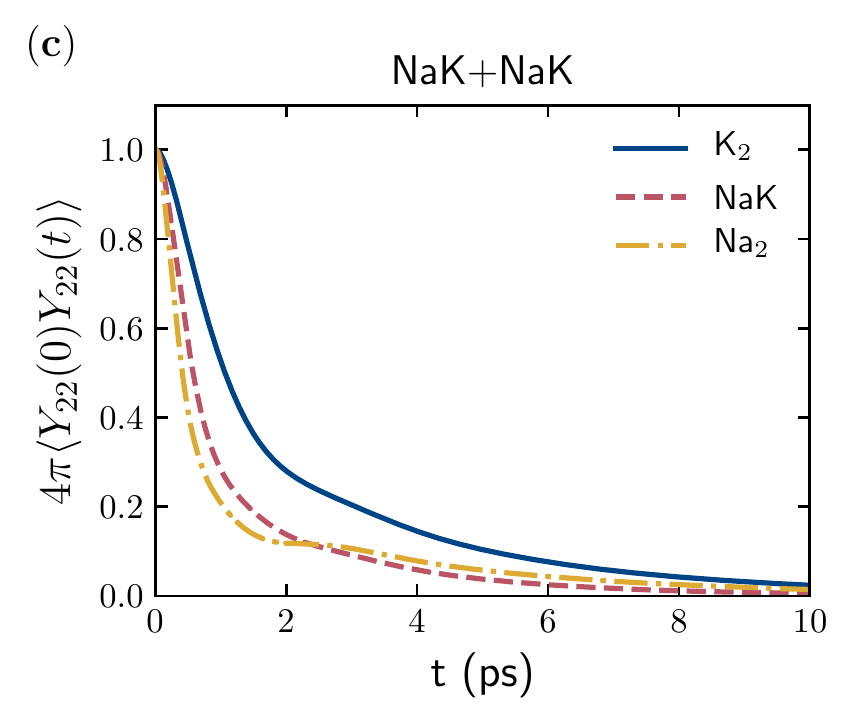}
    \includegraphics[width=0.475\textwidth]{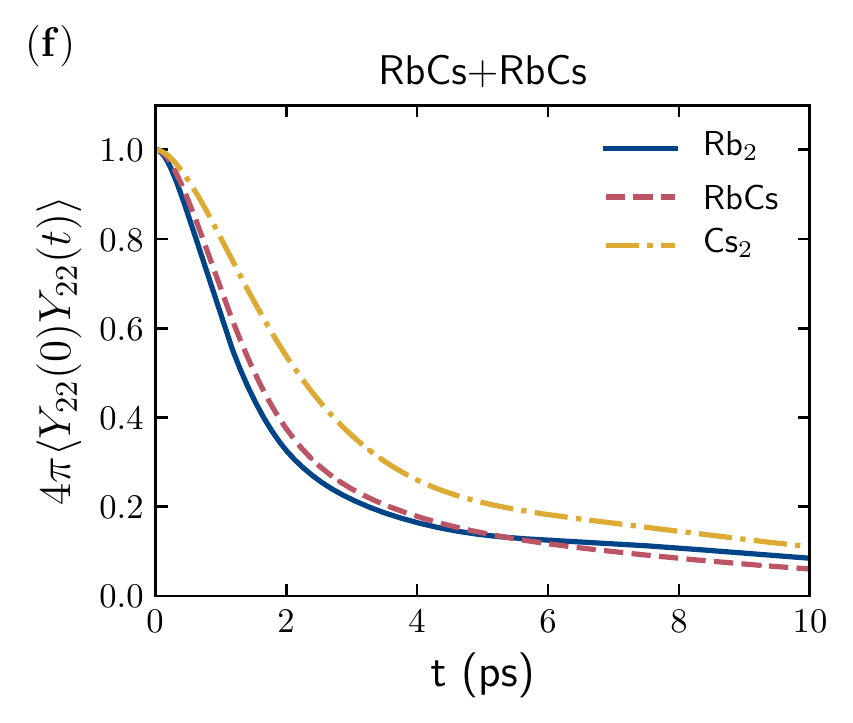}
    \caption{The autocorrelation of the spherical harmonics $Y_{2m}({\hat{r}}_{\text{AB}})$. }
    \label{fig:ACF_spherical}
\end{figure}

\begin{figure}
    \centering
    \includegraphics[width=0.475\textwidth]{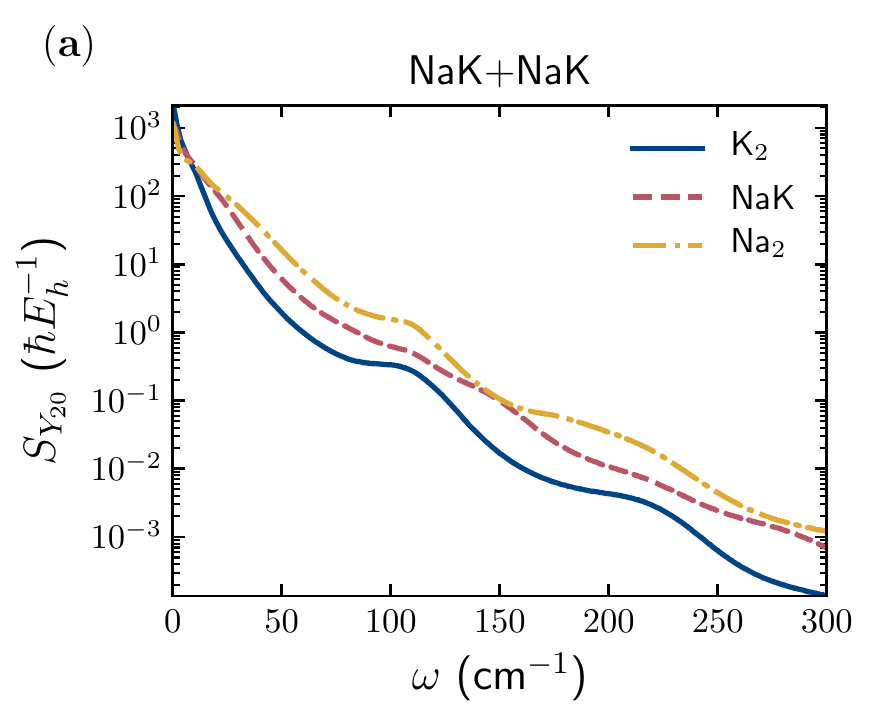}
    \includegraphics[width=0.475\textwidth]{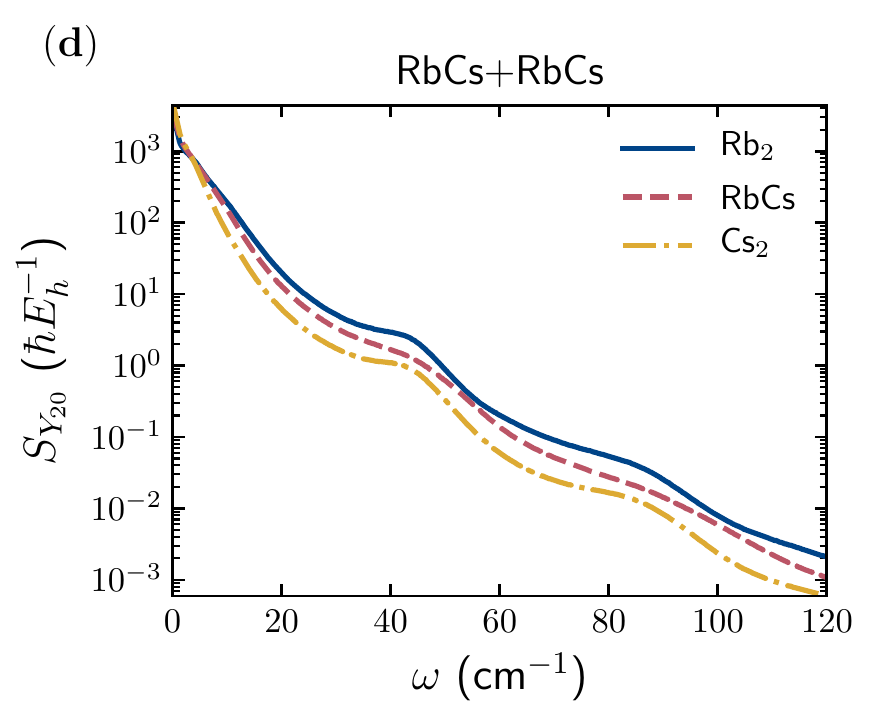}
    \includegraphics[width=0.475\textwidth]{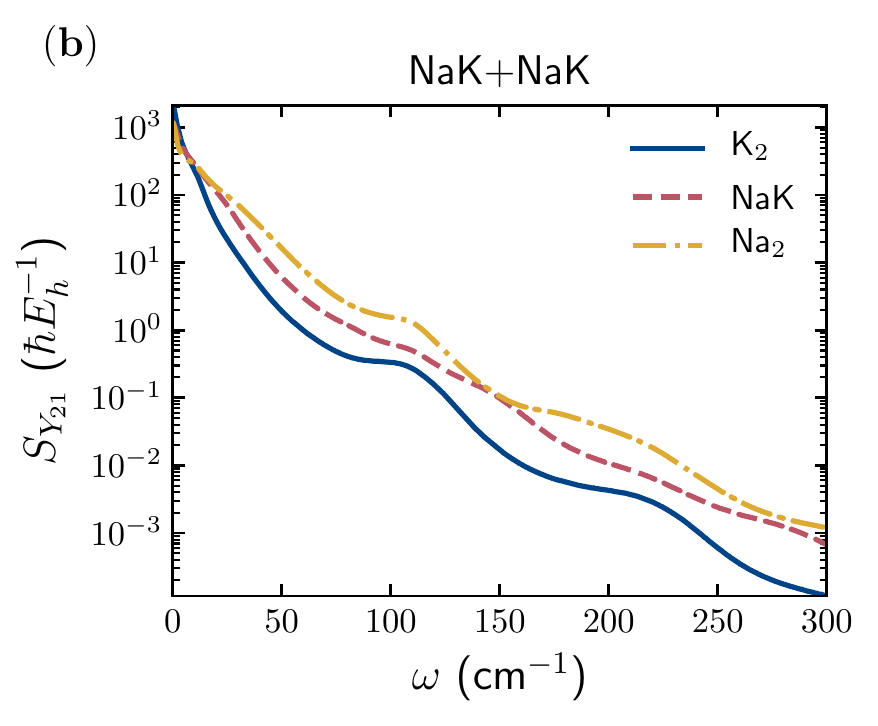}
    \includegraphics[width=0.475\textwidth]{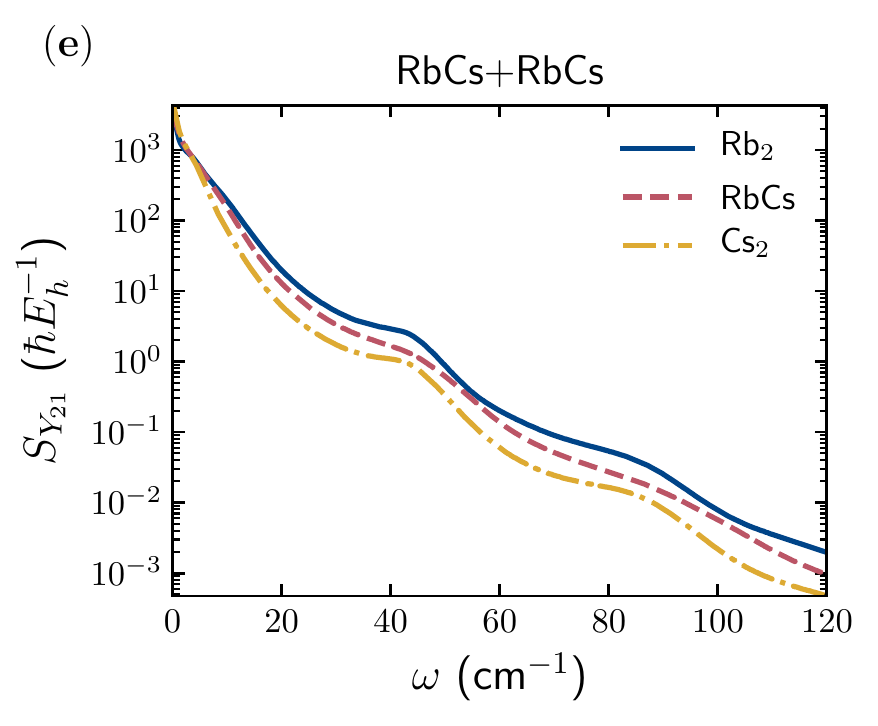}
    \includegraphics[width=0.475\textwidth]{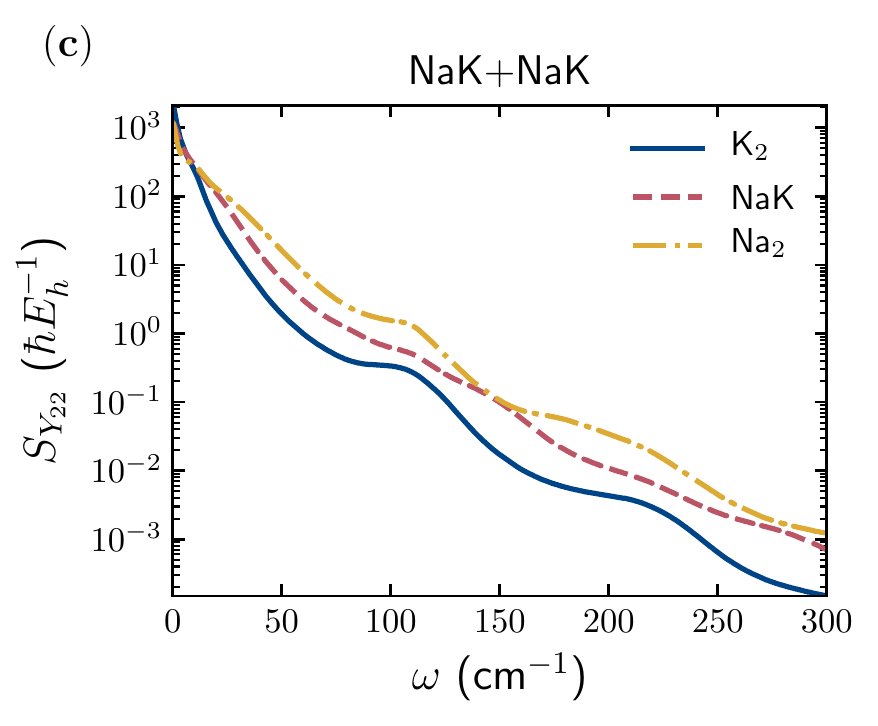}
    \includegraphics[width=0.475\textwidth]{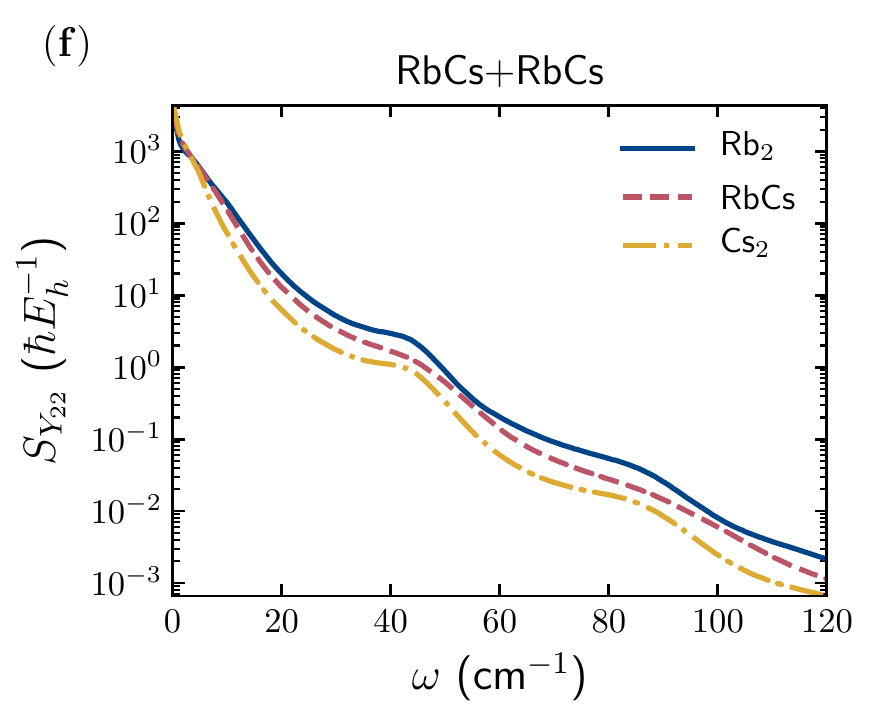}
    \caption{The Fourier transform of the autocorrelation of the spherical harmonics $S_{Y_{2m}({\hat{r}}_{\text{AB}})}(E_0,\omega)$.
    \label{fig:F_ACF_spherical_log}
}
\end{figure}

Using the same formalism as developed for static fields,
we can now determine the mean square coupling strength from the autocorrelation function of the coupling.
Apart from the constants specified in Eq.~\eqref{eq:hf_matel}, this amounts to computing the autocorrelation function of the second-rank spherical harmonics.
Unlike the dipole moment before, the spherical harmonics are complex valued and the required autocorrelation function becomes
\begin{align}
\left\langle Y_{2,m}\left(\hat{r}\left[t\right]\right) Y_{2,m}^{\ast} \left(\hat{r}\left[0\right]\right) \right\rangle &= \left\langle R_{2,m}\left(\hat{r}\left[t\right]\right) R_{2,m} \left(\hat{r}\left[0\right]\right) \right\rangle + \left\langle I_{2,m}\left(\hat{r}\left[t\right]\right) I_{2,m} \left(\hat{r}\left[0\right]\right) \right\rangle \nonumber \\
+& i \left\langle I_{2,m}\left(\hat{r}\left[t\right]\right) R_{2,m} \left(\hat{r}\left[0\right]\right) \right\rangle - i \left\langle R_{2,m}\left(\hat{r}\left[t\right]\right) I_{2,m} \left(\hat{r}\left[0\right]\right) \right\rangle \nonumber \\
&= \left\langle R_{2,m}\left(\hat{r}\left[t\right]\right) R_{2,m} \left(\hat{r}\left[0\right]\right) \right\rangle + \left\langle I_{2,m}\left(\hat{r}\left[t\right]\right) I_{2,m} \left(\hat{r}\left[0\right]\right) \right\rangle,
\end{align}
where ${R}_{2,m}$ and ${I}_{2,m}$ indicate the real and imaginary part of ${Y}_{2,m}$, respectively.
In the last step above we used that ${R}_{2,m}$ and ${I}_{2,m}$ separately are Hermitian operators,
and we used that the ensemble average is independent under time translation and the autocorrelation function symmetric under time reversal.
Hence, in practice, we need to compute only the sum of the autocorrelation functions of the real and imaginary parts of the spherical harmonics,
rather than the autocorrelation function of a complex-valued quantity.
Ultimately, we need the static limit of the Fourier transform of this autocorrelation function
\begin{align}
S_{Y_{2,m}} (E,\omega) = \frac{1}{2\pi} \int_{-\infty}^\infty \left\langle Y_{2,m}\left(\hat{r}\left[t\right]\right) Y_{2,m}^{\ast} \left(\hat{r}\left[0\right]\right) \right\rangle\ \exp(i \omega t) dt.
\end{align}
We calculated the autocorrelation and the Fourier transform of the autocorrelation, $S_{Y_{lm}({\hat{r}}_{\text{AB}})}(E_0,\omega)$, of the spherical harmonic functions $Y_{20}({\hat{r}}_{\text{AB}})$, $Y_{21}({\hat{r}}_{\text{AB}})$, and $Y_{22}({\hat{r}}_{\text{AB}})$ for all unique pairs of atoms $\text{AB}$ in the NaK+NaK and RbCs+RbCs systems using the same method as we used for the dipole autocorrelation function.
We show the autocorrelation functions in Fig.~\ref{fig:ACF_spherical}, and their Fourier transforms in Fig.~\ref{fig:F_ACF_spherical_log},
and we list their static limit in Table~\ref{tab:spherical_harmonic}.

Having obtained the autocorrelation function of the relevant spherical harmonics, we can compute the mean square coupling
between an initial state --- with initial spin state $m_i$ and mechanical total angular momentum $J=0$ --- and a final state --- with final spin state $m'_i$ and mechanical angular momentum $J'=2$ --- as
\begin{align}
\left\langle \left| \left\langle m_i, J=0 | \hat{H}_\mathrm{eQq}  | m_i', J'=2 \right\rangle \right|^2 \right\rangle = \frac{c_{i,m_i;i',m_i'}^2 S_{Y_{2,m'_i-m_i}} (E,0)}{\hbar 2} \left(1/\rho_0 + 1/\rho_2\right),
\label{eq:spincoup}
\end{align}
where $c_{i,m_i;i',m_i'}$ is the prefactor in the corresponding right-hand side of Eq.~\eqref{eq:hf_matel},
that is
\begin{align}
{c_{i,m_i;i',m_i'} = \langle i, m_i | \hat{H}_\mathrm{eQq} | i', m'_i\rangle / Y_{2,m'_i-m_i}(\hat{r})}.
\end{align}
Equation~\eqref{eq:spincoup} is the nuclear-spin-coupling equivalent of Eq.~\eqref{eq:var_autocorr} in the case of coupling by static electric fields.
As with static fields before, whether or not nuclear spin is conserved is determined by the dimensionless parameter $\Omega$, the mean coupling in units of the level spacing.
Explicitly, we have
\begin{align}
\Omega =  |c_{i,m_i;i',m_i'}| \sqrt{\frac{6^3}{10}} \sqrt{\rho_0 S_{Y_{2,m'_i-m_i}}(E,0)},
\end{align}
where we have assumed $\rho_J = (2J+1)\rho_0$.
This yields for both RbCs transitions considered here $\Omega = 0.06$ and for the NaK $\Delta m_i = 1$ and $2$ transitions $\Omega = 0.003$ and $0.001$, respectively.
These values of $\Omega$ are substantially smaller than unity indicating that nuclear spin is conserved in NaK+NaK and RbCs+RbCs collisions.
\begin{table}
    \centering
    \caption{Static limit of the spherical harmonics autocorrelation functions, $S_{Y_{lm}({\hat{r}}_{\text{AB}})}(E_0,0)$. }
    \begin{tabular}{cSSS}
    \hline
    \hline
         Diatom& \multicolumn{1}{c}{$S_{Y_{20}({\hat{r}}_{\text{AB}})}(E_0,0)$}& \multicolumn{1}{c}{$S_{Y_{21}({\hat{r}}_{\text{AB}})}(E_0,0)$}&\multicolumn{1}{c}{$ S_{Y_{22}({\hat{r}}_{\text{AB}})}(E_0,0)$} \\
         &\multicolumn{1}{c}{($\hbar E_h^{-1}$)}&\multicolumn{1}{c}{($\hbar E_h^{-1}$)}&\multicolumn{1}{c}{($\hbar E_h^{-1}$)}\\
         \hline
         \multicolumn{4}{c}{RbCs+RbCs: }\\
         RbCs&2970&2932&2949\\
         Rb$_2$&3272&3193&3226\\
         Cs$_2$&4374&4267&4306\\
         \multicolumn{4}{c}{NaK+NaK: }\\
         NaK&1150&1153&1144\\
         K$_2$&2089&2068&2070\\
         Na$_2$&1141&1145&1152\\
         \hline
         \hline
         
    \end{tabular}
   
    \label{tab:spherical_harmonic}
\end{table}

\section{The diatomics-in-molecules potentials}\label{sec:DIM}

In this section we describe in detail the model of the interaction potentials used throughout this work.
In absence of accurate \emph{ab initio} interaction potentials for the collision complexes we model the interactions using pairwise interactions between the constituent atoms using the ideas of diatomics-in-molecules \cite{ellison:63}.
That is, the interactions are pairwise atom-atom interactions, but we also account for the spin-dependence of this interaction.
Alkali metal atoms are in a $^2S$ electronic state, so an atom-atom pair can form either a spin singlet or spin triplet state.
The singlet potential is typically more strongly bound.
However, the ground state potential is not simply the sum of singlet interactions between all atoms,
as one atom cannot simultaneously spin-pair with multiple other atoms.
We use the method of diatomics-in-molecules \cite{ellison:63} to account for this as follows.

The spin dependence of the interaction between atoms $1$ and $2$ at bond length $r_{1,2}$ is written as
\begin{align}
\hat{V}^{(1,2)}(r_{1,2}) = V_\mathrm{singlet}(r_{1,2}) + \frac{1}{2} \hat{s}_{1,2}^2 \left[V_\mathrm{triplet}\left(r_{1,2}\right)-V_\mathrm{singlet}\left(r_{1,2}\right)\right],
\end{align}
where $\hat{s}_{a,b}^2 = \hat{\vec{s}}_{(a,b)}\cdot\hat{\vec{s}}_{(a,b)}$ and $\hat{\vec{s}}_{(a,b)} = \hat{\vec{s}}_{(a)} + \hat{\vec{s}}_{(b)}$ is the vector sum of the electronic spin operator for atoms $a$ and $b$.
For a given configuration of nuclei we compute all the interatomic distances,
and form the total interaction as the sum of pair interactions between all constituent atoms, $\hat{V} = \hat{V}^{(1,2)}(r_{1,2}) + \hat{V}^{(1,3)}(r_{1,3}) + \hat{V}^{(2,3)}(r_{2,3})$ for atom-diatom collision complexes, and similar for the diatom-diatom case.

To calculate the lowest adiabatic potential, we first compute a matrix representation of $\hat{V}$ in a basis of coupled spin functions.
For the atom-diatom case we use the basis
\begin{align}
|s_{1,2}; S M_S\rangle = \sum_{m_1,m_2,m_3} |s_1 m_1\rangle|s_2 m_2\rangle|s_3 m_3\rangle \langle s_1 m_1 s_2 m_2 | s_{1,2}, m_{1}+m_2 \rangle \langle s_{1,2}, m_{1}+m_2, s_3, m_3 | S M_S \rangle,
\end{align}
where $\langle s_1 m_1 s_2 m_2 | s_3 m_3\rangle$ is a Clebsch-Gordan coefficient.
For the diatom-diatom case
\begin{align}
|s_{1,2}, s_{3,4}; S M_S \rangle &= \sum_{m_1,m_2,m_3,m_4} |s_1 m_1\rangle|s_2 m_2\rangle|s_3 m_3\rangle|s_4 m_4\rangle \langle s_1 m_1 s_2 m_2 | s_{1,2}, m_{1}+m_2 \rangle \nonumber \\
\times& \langle s_3 m_3 s_4 m_4 | s_{3,4}, m_{3}+m_4 \rangle \langle s_{1,2}, m_1+m_2, s_{3,4}, m_3+m_4 | S M_S\rangle.
\end{align}
Because all atoms are in a $^2S$ state, $s_1=s_2=s_3=s_4={}^1/_{2}$.
The total spin quantum numbers $S$ and $M_S$ are good quantum numbers.
For the atom-diatom and diatom-diatom collision complexes $S=1/2$ and $S=0$, respectively.
The matrix representation of the electronic interaction $\hat{V}$ is independent of $M_S$.

In both atom-diatom and diatom-diatom cases, we obtain a two-dimensional basis set.
For the atom-diatom case we have $\{|s_{1,2}; S, M_S\rangle = |0;\ {}^1/_2, M_S\rangle, |1;\ {}^1/_2, M_S\rangle\}$,
and the relevant matrix representations of the spin operators are given by
\begin{align}
\mathbf{s}_{1,2}^2 = \begin{bmatrix} 0 & 0\\ 0& 2 \end{bmatrix},
\qquad
\mathbf{s}_{1,3}^2 = \begin{bmatrix} \frac{3}{2} & -\frac{1}{2} \sqrt{3} \\ -\frac{1}{2} \sqrt{3} & \frac{1}{2} \end{bmatrix},
\qquad
\mathbf{s}_{2,3}^2 = \begin{bmatrix}\frac{3}{2} & \frac{1}{2} \sqrt{3} \\ \frac{1}{2} \sqrt{3} & \frac{1}{2} \end{bmatrix}.
\end{align}

For the diatom-diatom case, we use the basis $\{|s_{1,2}, s_{3,4}; S M_S \rangle = |0, 0; 0 0\rangle, |1, 1; 0 0\rangle\}$ and the relevant matrix representations of the spin operators are given by
\begin{align}
\mathbf{s}_{1,2}^2 = \begin{bmatrix} 0 & 0\\ 0& 2 \end{bmatrix},
\qquad
\mathbf{s}_{1,3}^2 = \begin{bmatrix} \frac{3}{2} & -\frac{1}{2} \sqrt{3} \\ -\frac{1}{2} \sqrt{3} & \frac{1}{2} \end{bmatrix},
\qquad
\mathbf{s}_{1,4}^2 = \begin{bmatrix} \frac{3}{2} & \frac{1}{2} \sqrt{3} \\ \frac{1}{2} \sqrt{3} & \frac{1}{2} \end{bmatrix}, \nonumber \\
\mathbf{s}_{2,3}^2 = \begin{bmatrix} \frac{3}{2} & \frac{1}{2} \sqrt{3} \\ \frac{1}{2} \sqrt{3} & \frac{1}{2} \end{bmatrix},
\qquad
\mathbf{s}_{2,4}^2 = \begin{bmatrix} \frac{3}{2} & -\frac{1}{2} \sqrt{3} \\ -\frac{1}{2} \sqrt{3} & \frac{1}{2} \end{bmatrix},
\qquad
\mathbf{s}_{3,4}^2 = \begin{bmatrix} 0 & 0\\ 0& 2 \end{bmatrix}.
\end{align}

Using the matrix representations given above one can easily set up a $2\times 2$ matrix representation of the interaction operator,
and obtain the diatomics-in-molecules interaction potential as the lowest eigenvalue.

The interaction potential displayed as the surface plot in Fig.~1(a) of the main text,
is obtained as a two-dimensional cut of the NaK+K interaction potential with the NaK bond length fixed it its equilibrium position.
This is included only as an illustration.

\end{document}